\documentclass[twocolumn]{aastex62}

\usepackage{amsmath}
\usepackage{longtable}
\usepackage{lipsum}

\newcommand{\feii}{Fe\,{\sc ii}}
\newcommand{\oiii}{{\sc [}O\,{\sc iii]}}

\newcommand{\mgii}{Mg\,{\sc ii}}
\newcommand{\ha}{H$\alpha$}
\newcommand{\hb}{H$\beta$}

\shorttitle{Quasar Host Galaxies at $z<0.8$}
\shortauthors{Yue et al.}

\begin{document}

\title{The Sloan Digital Sky Survey Reverberation Mapping Project: Quasar Host Galaxies at $z<0.8$ from Image Decomposition}

\correspondingauthor{Minghao Yue}
\email{yuemh@email.arizona.edu}

\correspondingauthor{Linhua Jiang}
\email{jiangKIAA@pku.edu.cn}

\author{Minghao Yue}
\affiliation{Kavli Institute for Astronomy and Astrophysics, Peking University, Beijing 100871, China}
\affiliation{Steward Observatory, University of Arizona, 933 North Cherry Avenue, Tucson, AZ 85719, USA}

\author{Linhua Jiang}
\affiliation{Kavli Institute for Astronomy and Astrophysics, Peking University, Beijing 100871, China}

\author{Yue Shen}
\affiliation{Department of Astronomy, University of Illinois at Urbana-Champaign, Urbana, IL 61801, USA}
\affiliation{National Center for Supercomputing Applications, University of Illinois at Urbana-Champaign, Urbana, IL 61801, USA}
\affiliation{Alfred P. Sloan Research Fellow}

\author{Patrick B. Hall}
\affiliation{Department of Physics and Astronomy, York University, Toronto, ON M3J 1P3, Canada}

\author{Zhefu Yu}
\affiliation{Kavli Institute for Astronomy and Astrophysics, Peking University, Beijing 100871, China}
\affiliation{Department of Astronomy, The Ohio State University, 140 West 18th Ave., Columbus, OH 43210, USA}

\author{Donald P. Schneider}
\affiliation{Department of Astronomy and Astrophysics, The Pennsylvania State University, 525 Davey Laboratory, University Park, PA 16802, USA}
\affiliation{Institute for Gravitation and the Cosmos, The Pennsylvania State University, University Park, PA 16802, USA}

\author{Luis C. Ho}
\affiliation{Kavli Institute for Astronomy and Astrophysics, Peking University, Beijing 100871, China}
\affiliation{Department of Astronomy, School of Physics, Peking University, Beijing 100871, China}

\author{Keith Horne}
\affiliation{SUPA Physics and Astronomy, University of St. Andrews, KY16 9SS, Scotland, UK}

\author{Patrick Petitjean}
\affiliation{Institut d'Astrophysique de Paris, CNRS$-$Universit\'e Pierre et Marie Curie, 98bis boulevard Arago, 75014 Paris, France}

\author{Jonathan R. Trump}
\affiliation{Department of Physics, University of Connecticut, 2152 Hillside Road, Storrs, CT 06269}



\begin{abstract}

We present the rest-frame UV and optical photometry and morphology of low-redshift 
broad-line quasar host galaxies from the Sloan Digital Sky Survey Reverberation Mapping 
project. Our sample consists of 103 quasars at $z<0.8$, spanning a 
luminosity range of $-25\le M_g\le -17$ mag. We stack the multi-epoch images in 
the $g$ and $i$ bands taken by the Canada-France-Hawaii Telescope. The 
combined $g$-band ($i$-band) images reach a $5\sigma$ depth of 26.2 (25.2) mag, 
with a typical PSF size of $0\farcs7$ ($0\farcs6$). 
Each quasar is decomposed into a PSF and a S\'ersic profile, representing the central
 AGN and the host galaxy components, respectively. The systematic errors of 
the measured host galaxy flux in the two bands are 0.23 and 0.18 mag. 
The relative errors of the measured galaxy half-light radii 
($R_e$) are about 13\%. We estimate the rest-frame $u$- and $g$-band flux 
of the host galaxies, and find that the AGN-to-galaxy flux ratios in the $g$ band 
are between 0.9 to 4.4 (68.3\% confidence). These galaxies have 
high stellar masses $M_\ast = 10^{10}\sim10^{11}\, M_\odot$.
They have similar color with star-forming galaxies
at similar redshifts, in consistent with AGN positive feedback
in these quasars.
We find that the $M_*-M_\text{BH}$ relation in our sample is 
shallower than the local $M_\text{Bulge}-M_\text{BH}$ relation. 
The S\'ersic indices and the $M_*-R_e$ relation indicate that the majority 
of the host galaxies are disk-like.

\end{abstract}


\section{Introduction} \label{intro}

AGNs are powerful objects where supermassive black holes (SMBHs) in galaxy centers are actively accreting materials, 
releasing huge amounts of energy by radiation and material outflows.
AGNs are believed to have strong impact on their host galaxies, known as AGN feedback
\citep[for recent reviews, see][]{fabian12,king15}. 
Such feedback,  including ``negative feedback'' and ``positive feedback'', can significantly influence AGN host galaxies in many aspects, especially star formation.
The negative feedback scenario suggests that jets and radiative winds from 
AGN quench star formation by heating and/or expelling cold gas in host galaxies.
This scenario provides a possible solution to many key questions in galaxy formation,
such as the different shapes between the mass functions of galaxies and dark matter halos at the high mass end.
It has been supported by some simulations \citep[e.g.,][]{DM05,Springel05,hopkins06,croton06}.
Observations have also found evidence for AGN-driven outflows \citep[e.g.,][]{cicone14}
and AGN-heated gas around massive quiescent elliptical galaxies \citep[e.g.,][]{spacek16}.
On the contrary, other simulations have shown that the outflow jets may disturb gas in host galaxies,
enhancing star formation \citep[e.g.,][]{zinn13}, which suggests AGN positive feedback.
Supporting evidence includes observations that 
star-forming regions in AGN host galaxies have a significant alignment with jets \citep[e.g.,][]{salome15}.

To determine which mechanism dominates, there have been efforts to 
measure star formation rates (SFRs) and stellar populations in AGN host galaxies.
Results from early studies were controversial. For example, \citet{Kirhakos99} claimed that quasar host galaxies 
had bluer colors than normal galaxies, suggesting active star formation. \citet{mclure99} 
showed that quasar host galaxies had old stellar populations, indicating low recent SFRs.
AGN feedback is likely a mixture of positive and negative feedback \citep[e.g.,][]{zinn13}, 
and the feedback process can be dominated by either of the mechanisms.
In addition, the properties of quasar host galaxies may also depend on redshift and luminosity, 
which makes the situation more complex.

Large-area sky surveys, such as the Sloan Digital Sky Survey \citep[SDSS;][]{york00}, have  
significantly contributed to the study of quasar/AGN host galaxy properties in the past two decades.
A commonly used method to study the AGN impact is to analyze the stellar populations of host galaxies.
Several recent studies suggest that the host galaxies of unobscured broad-line AGN 
are massive and systematically bluer than normal galaxies \citep[e.g.,][]{jahnke04,trump13},
although this result may be largely due to sample selection effects \citep[e.g.,][]{aird12}.
It has also been recognized that AGN feedback may strongly depend on many properties of AGNs.
For example, \citet{Kauffmann03} studied type-II AGNs from SDSS and found that these AGNs were almost exclusively 
hosted by massive galaxies with stellar mass $M_*>10^{10}M_{\odot}$. They also reported that the host galaxies 
of low-luminosity type-II AGNs had stellar populations similar to early type galaxies, 
while the host galaxies of high-luminosity AGNs had much younger stellar populations.
\citet{Hickox09} examined a sample of 585 AGNs and concluded that the hosts of radio AGNs were located in ``the red sequence", 
X-ray selected AGNs were located in ``the green valley", and infrared selected AGNs were bluer than X-ray selected quasars.
The dependence of host galaxy properties on AGN types and properties indicates that it is necessary to have thorough studies on
all types of AGNs.

For the most luminous AGNs, i.e., unobscured (Type-I) quasars, the measurement of host galaxies is difficult,
and often subject to large uncertainties due to the contamination from the quasar light.
Currently there are three techniques that are widely used to study quasar host galaxies: 
spectra energy distribution (SED) fitting, image decomposition, and spectra decomposition.
Unlike SED fitting and spectra decomposition, image decomposition does not depend on spectra/SED models of quasars and galaxies. 
The only major assumption is that the quasar component can be modeled as a point spread function (PSF). 
Image decomposition can provide the morphological information of host galaxies 
which can be used to constrain quasar triggering models \citep[e.g.,][]{cisternas11,villforth17}.

Early studies of AGN image decomposition mainly used Hubble Space Telescope images
\citep[e.g.,][]{bahcall97,Kirhakos99,jahnke04,kim08,villforth17}. These samples were usually small.
Image decomposition studies using ground-based data 
can have samples of several hundred of quasars. For example,
\citet{mat14} performed image decomposition for a sample of $\sim800$ quasars at $z<0.6$ from the SDSS Stripe 82.
The typical PSF size of their images is about $1\farcs0-1\farcs1$.
They suggested that quasar host galaxies are systematically bluer than normal galaxies.
Meanwhile, 
the systematic errors introduced by the decomposition procedure are poorly understood.
For example, \citet{bettoni15} fitted the Stripe 82 images of low-redshift SDSS quasars using a different method,
and found that quasar host galaxies have similar colors compared to a redshift-matched sample of inactive galaxies, 
in contrary to the results of \citet{mat14}.

In order to obtain reliable measurements on quasar host galaxies, high-quality images are needed.
In this work, we use deep images from the SDSS Reverberation Mapping (SDSS-RM) project 
to study 103 quasar host galaxies at $z<0.8$. Our combined $i$-band images reach a $5\sigma$ depth of $>25$ mag 
with a PSF FWHM of $\sim 0\farcs6$. The depth and PSF of our images,
two crucial factors for the image decomposition analysis, 
are significantly better than those of the ground-based images in most previous studies.
Our paper is organized as follows. Section \ref{data} describes the imaging and spectral data, and the quasar sample in our work. 
Section \ref{image} presents our image decomposition method. 
A spectroscopic analysis method that makes use of the result from the image decomposition is discussed in Section \ref{spectra}. 
Section \ref{results} presents the results, Section \ref{discussion} presents some further discussions, and Section \ref{sum} summarizes this paper. 
We use a $\Lambda$-dominated flat cosmology with $H_0=70$ km s$^{-1}$ Mpc$^{-1}$, $\Omega_{m}=0.3$, and $\Omega_{\Lambda}=0.7$. 
We use AB magnitude \citep{oke83} through this paper.

\section{Data and Quasar Sample} \label{data}

\begin{table*}
\caption{Co-added Imaging Data}
\hspace*{-2.7cm}
\begin{tabular}{c|ccccccccc}
\hline\hline
Pointing & A & B & C & D & E & F & G & H & I\\\hline
R.A. &14$^\text{h}14^\text{m}51^\text{s}$ &14$^\text{h}14^\text{m}51^\text{s}$ &14$^\text{h}08^\text{m}47^\text{s}$ &14$^\text{h}08^\text{m}19^\text{s}$ &14$^\text{h}08^\text{m}39^\text{s}$ &14$^\text{h}14^\text{m}52^\text{s}$ &14$^\text{h}21^\text{m}03^\text{s}$ &14$^\text{h}21^\text{m}24^\text{s}$ &14$^\text{h}20^\text{m}54^\text{s}$\\
Decl. &52$^\circ$05$'$28$''$ &52$^\circ$06$'$35$''$ &52$^\circ$09$'$27$''$ &53$^\circ$05$'$13$''$ &54$^\circ$01$'$20$''$ &54$^\circ$04$'$09$''$ &54$^\circ$01$'$35$''$ &53$^\circ$05$'$30$''$ &52$^\circ$09$'$28$''$\\
$\text{N}_{\text{image},g}$&157&101&97&97&92&91&91&94&99\\
$\text{N}_{\text{image},i}$&114&74&70&70&68&69&70&70&70\\
PSF FWHM ($g$)($''$)&0.69&0.72&0.72&0.72&0.73&0.72&0.70&0.71&0.72\\
PSF FWHM ($i$)($''$)&0.55&0.56&0.57&0.56&0.58&0.57&0.57&0.56&0.57\\
Mag Limit ($g$)&26.4&26.1&26.2&26.2&26.1&26.1&26.2&26.2&26.2\\
Mag Limit ($i$)&25.4&25.2&25.3&25.3&25.2&25.2&25.2&25.3&25.2\\\hline
\end{tabular}
		\vspace{0.2cm}
		\centering {\bf{Notes.}} All magnitude limits are $5\sigma$ for point sources.
\label{tbl:pointing}
\end{table*}

\subsection{Imaging and Spectroscopic Data}
In this study, we use the optical images and spectra from the SDSS-RM project to analyze quasar host galaxies.
We first decompose the $g$- and $i$-band images of each quasar into a PSF component and a S\'ersic profile component.
Based on the flux ratio of the two components, the spectrum of a quasar is decomposed into
an AGN component and a galaxy component. The AGN component is described as the combination of 
a power-law continuum and emission lines.
The rest-frame flux of the host galaxy is then calculated using the galaxy component of the spectrum.
We apply this method to analyze host galaxy properties, rather than simply adopting the flux from the image decomposition,
because we do not have enough bands to perform the traditional $k$-correction.
We will describe the details later.

As part of the SDSS-III program \citep{eisenstein11}, SDSS-RM is a multi-object reverberation mapping project, 
monitoring 849 broad-line quasars in a 7 deg$^2$ field. 
It aims to detect the time lag between the continuum and the broad line region variabilities of quasars, 
using both spectroscopic and photometric observations. 
In this study, we use the co-added optical images and spectra.
The spectroscopy was made by the Baryonic Oscillation Spectroscopic Survey (BOSS)
spectrograph mounted on the SDSS 2.5m telescope \citep{gun06}, which provides a wavelength coverage 
from 3650 to 10,500 {\AA} and a resolution $R\sim2000$ \citep{smee13}. 
The photometric monitoring of SDSS-RM was done at 
the Steward Observatory Bok telescope,
the Kitt Peak National Observatory (KPNO) 4m telescope,
and the Canada-France-Hawaii Telescope (CFHT).
The observations were conducted in 2014, with a cadence of about 2 days 
in the $g$ and $i$ bands.

In this work, we use images taken by the CFHT using the MegaCam instrument that
consists of 36 CCD chips with a pixel scale of $0\farcs187$ \citep{aune03}.
The CFHT images have excellent PSFs ($\sim 0\farcs6$ in the $i$ band), which are 
much better than the images taken by the other two telescopes. 
To cover the entire SDSS-RM field,
a total of 9 pointings were used (denoted as points A to I; Table \ref{tbl:pointing}), 
and the images at each pointing consist of two dither positions to cover CCD gaps.
The detailed information about the observations can be found in \citet{shen15}.
There are 1067 images in the $g$ band and 794 images in the $i$ band.
The typical integration time per exposure is 78 s in $g$ and 111 s in $i$.

\subsection{Image Co-addition}\label{coadd}

In this section, we present our image co-addition method. 
We first reject images that have poor quality recorded in the observation logs.
We further remove cosmic rays from the images using the {\texttt{LA-Cosmic}} algorithm \citep{vdokkum01}.

\subsubsection{Image Selection and Co-addition} \label{imgselect}
For each image, we first estimate three parameters: atmospheric extinction (or sky transparency), 
PSF FWHM, and sky background. We run {\texttt{SExtractor}} \citep{bertin96}, and select bright and isolated 
point sources. The transparency and PSF FWHM are estimated from the photometry and FWHM values of these objects.
The sky background is the median value of the image. 
We then reject images with PSF FWHM values among the largest 10\%, images with 
sky background among the largest 5\%, and images with atmospheric extinction among the largest 5\%. 
The typical number of the remaining images at one pointing 
 is $\sim 100$ in the $g$ band and $\sim 70$ in the $i$ band.
We utilize a ``weighted average'' co-addition. 
Following the method used for the SDSS Stripe 82 image co-addition \citep{annis14, jiang14}, 
each image is assigned a weight proportional to $T/(\text{FWHM}^2\sigma^2)$,
where $T$ is the sky transparency, FWHM is the PSF FWHM, and $\sigma$ is the background 
noise. 
Since the background noise is dominated by the Poisson noise of sky background in our images, 
we assume $\sigma^2$ is proportional to sky background. 
We use {\texttt{SWarp}} \citep{bert02} to perform the co-addition. 

\subsubsection{Quality of Co-added Images}
Image decomposition of quasar host galaxies requires high image quality. 
Our co-added images have great depth and PSF compared to the ground-based images in previous studies.
The typical $5\sigma$ depth is $26.2$ mag in $g$ and $25.2$ mag in $i$ for point sources. 
They are about one magnitude deeper
than the combined SDSS Stripe 82 images.
The PSF FWHM values of our images are about $0\farcs7$ and $0\farcs6$ in the $g$ and $i$ bands, respectively. 
The variation of the PSF FWHM across an image is small. 
Over the entire SDSS-RM field, the variation is less than 15\%. 
Note that images with the largest 5\%\ PSFs have been removed earlier.
The PSF variation will be taken into account 
in the image decomposition process. More
information about co-added images is listed in Table \ref{tbl:pointing}.

\subsection{Quasar Sample}

\begin{figure}
\epsscale{1.2}
\includegraphics[trim={1cm 0 0 0},width=0.5\textwidth]{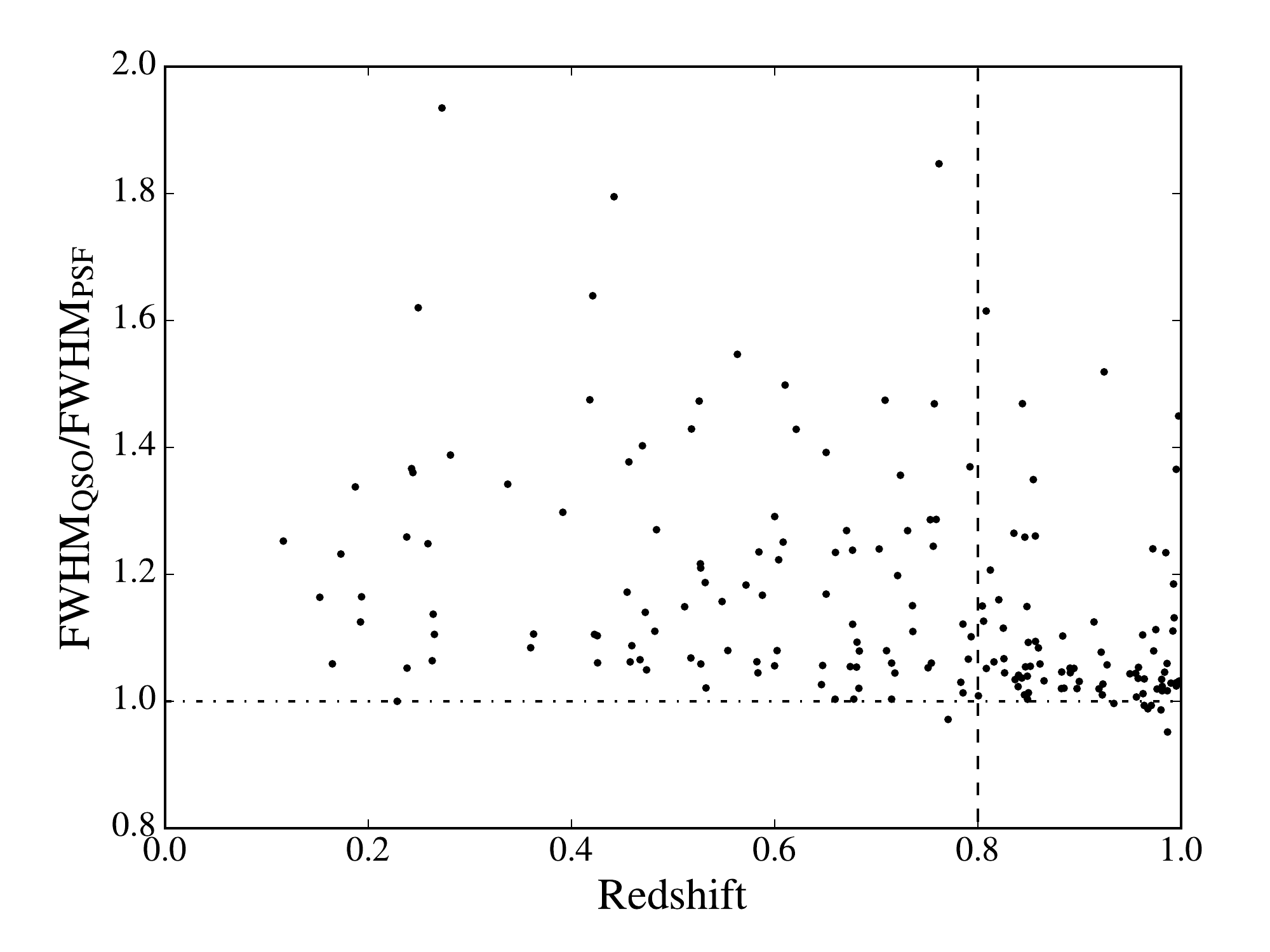}
\caption{The distribution of the quasar redshifts and FWHM. 
		Quasars with $\text{FWHM}_\text{QSO}/\text{FWHM}_\text{PSF}>1$ 
		are likely to have resolved host galaxy components. 
		The dashed line indicates our redshift cut $z<0.8$.}
\label{fig:select}
\end{figure}

\begin{figure}
\epsscale{1.25}
\includegraphics[trim={1.5cm 0 1cm 0},width=0.48\textwidth]{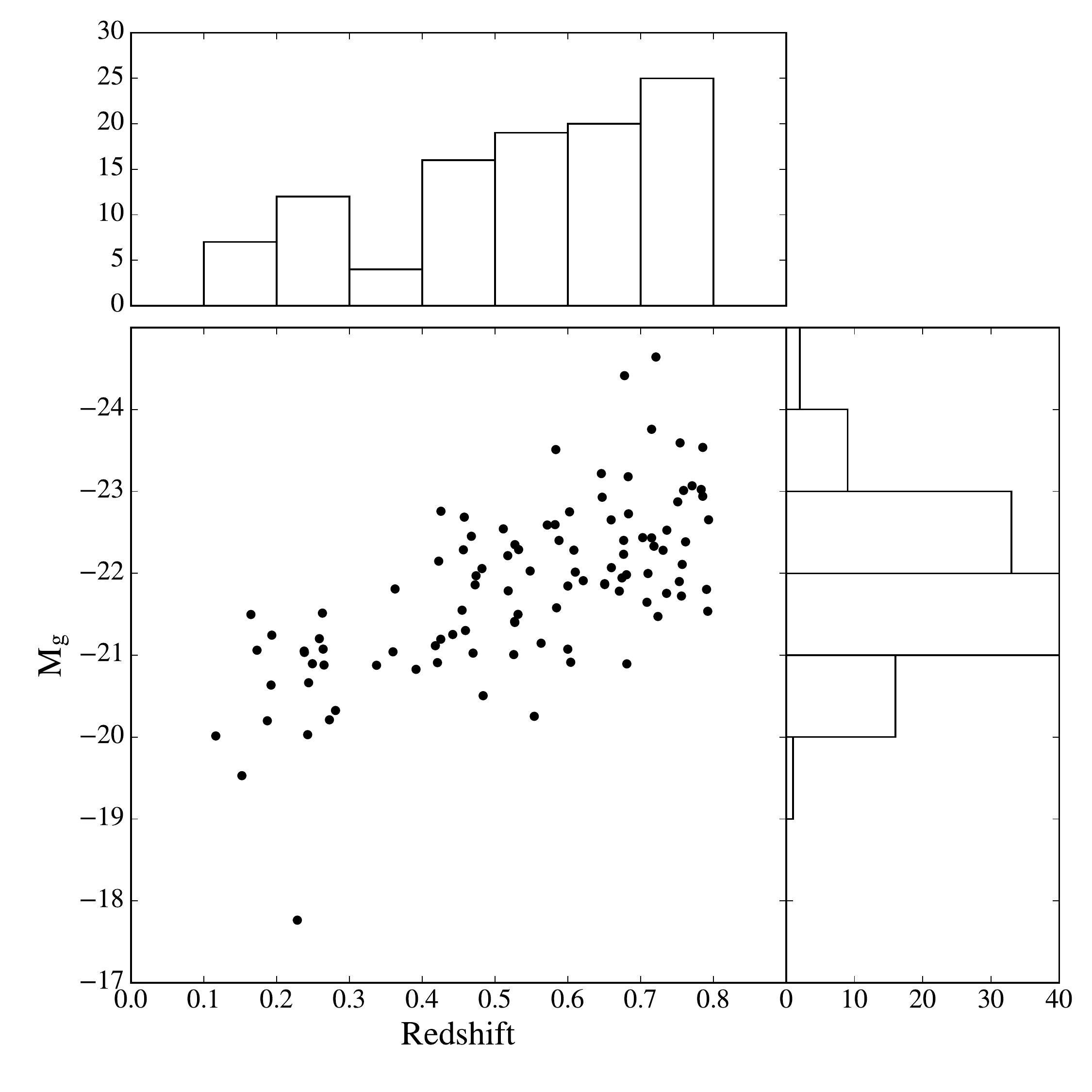}
\caption{The distribution of the redshifts and the $g$-band absolute magnitudes $M_{g}$ of the 103 quasars in our sample.}
\label{fig:zmi}
\end{figure}

The SDSS-RM quasar sample consists of 849 quasars with $i<21.7$ mag.
To estimate the redshift range in which quasars are resolved in our images,
we examine the relation between the redshifts and the FWHM of the quasar images.
Figure \ref{fig:select} shows the relation in the $i$ band. 
Most quasars at $z<0.8$ have FWHM larger than the PSF FWHM in both $g$ and $i$ bands, 
so we select quasars at
$z<0.8$ to construct our sample. This ensures a high success rate for image decomposition.
There are a total of 105 $z<0.8$ quasars in the SDSS-RM sample. 
We visually inspect all these quasars, and exclude two quasars
that are blended with nearby objects, 
or located at image edges. Our final sample consists of 103 quasars at $z<0.8$.
The distributions of their redshifts and $g$-band absolute magnitudes are shown in Figure \ref{fig:zmi}. 
More than half of the quasars are at $z>0.5$.


\section{Image Decomposition and Simulations} \label{image}
 
\subsection{Image Analysis} 

\label{imganalysis} 

In the following text, we use ``AGN component'' and ``galaxy component'' to denote
the central AGN (point source) and the host galaxy, respectively. Meanwhile, a ``quasar'' refers to the whole system, 
including the AGN and its host galaxy.

For each quasar, we decompose its image into a PSF 
(the AGN component) and a S\'ersic profile (the galaxy component). 
Our procedure is similar to \citet{mat14}.
We first resample the image so that the quasar center is located in the center of
a pixel. The pixel scale is reserved.
A local background is measured and subtracted. 
The PSF map of each image is modeled by {\texttt{PSFEx}} \citep{bertin11}. 
{\texttt{PSFEx}} selects bright point sources according to their half-light radii and flux, 
and fits PSFs to these sources. 
The output of {\texttt{PSFEx}} is a PSF map of a polynomial function of positions. 
More information about PSF modeling can be found in Appendix \ref{ap:psf}.
The PSF of a quasar is determined based on its position in the image. 
The PSF component has one free parameter, its flux. 
The S\'ersic function \citep{Sersic68} describes the radial profile of a galaxy and 
has the form 
\begin{equation}
I(r)=I_e\times \text{exp}\{-b_n[(\frac{r}{R_e})^{\frac{1}{n}}-1]\},
\end{equation}
where $n$ is the S\'ersic index that determines the shape of the profile, 
and $R_e$ is the effective radius that includes half of the total galaxy flux. 
By this definition, we can determine $b_n$ as a function of $n$. 
Therefore, a S\'ersic profile has three free parameters: $I_e$, $R_e$ and $n$.
A Gaussian profile has $n=1/2$, an exponential disk has $n=1$,
and a de Vaucouleurs profile has $n=4$. Most galaxies have $0.5<n<5$. 
The S\'ersic profile is convolved with the PSF to model the galaxy image.

We fit the 1-dimensional (1-D) radial profile of each quasar. 
The radial profile $I(j)$ at the $j$th data point 
is the mean value of all pixels 
whose distances to the object center $r$ satisfy $j-1<r\leq j$.
Data points with $1 \leq j \leq 10$ in the 1-D profiles are fitted. 
The central pixel $j=0$ is excluded in the fitting process, because the error of the PSF model 
is usually large in the center. 
We fit the image by minimizing the $\chi^2$ value defined as 
\begin{equation}
\chi^2=\sum_j \frac{[I(j)-I_{P}P(j)-I_{G}G(j,R_e,n)]^2}{\sigma_j^2},
\end{equation}
where $P(j)$ is the 1-D profile of the flux-normalized PSF, 
$G(j,R_e,n)$ is the 1-D profile of the flux-normalized model galaxy image, 
$I_P$ and $I_G$ are the intensities of PSF and galaxy components, 
and $\sigma_j$ is the uncertainty of the $j$th data point. 
The uncertainty is calculated by $\sigma_j^2=\sigma_{b,j}^2+\text{DN}_j/gain$, 
where $\sigma_{b,j}$ is the background noise at pixel $j$, 
DN$_j$ is the digital number of pixel $j$, and $gain$ is the gain of the image in e$^-$/ADU. 

\begin{figure*}
\centering
\includegraphics[width=0.9\textwidth]{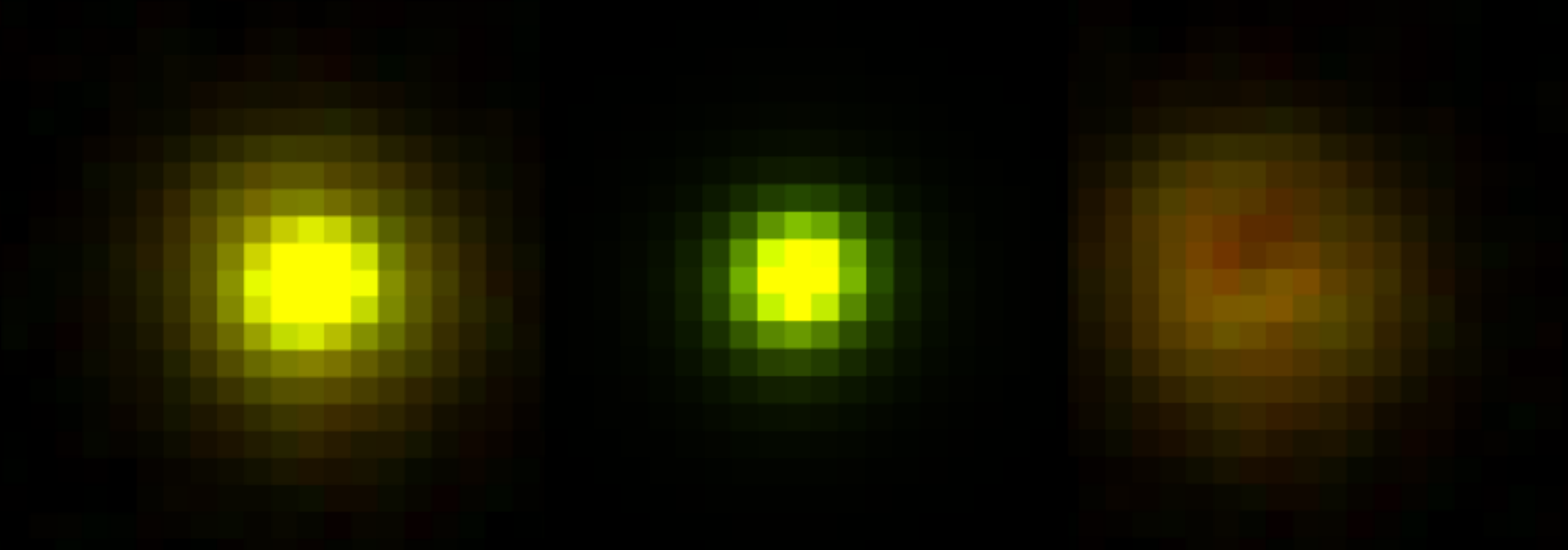}
\begin{minipage}{0.49\textwidth}
\raggedleft
\includegraphics[trim={0 0 0.6cm 0},width=1.0\textwidth]{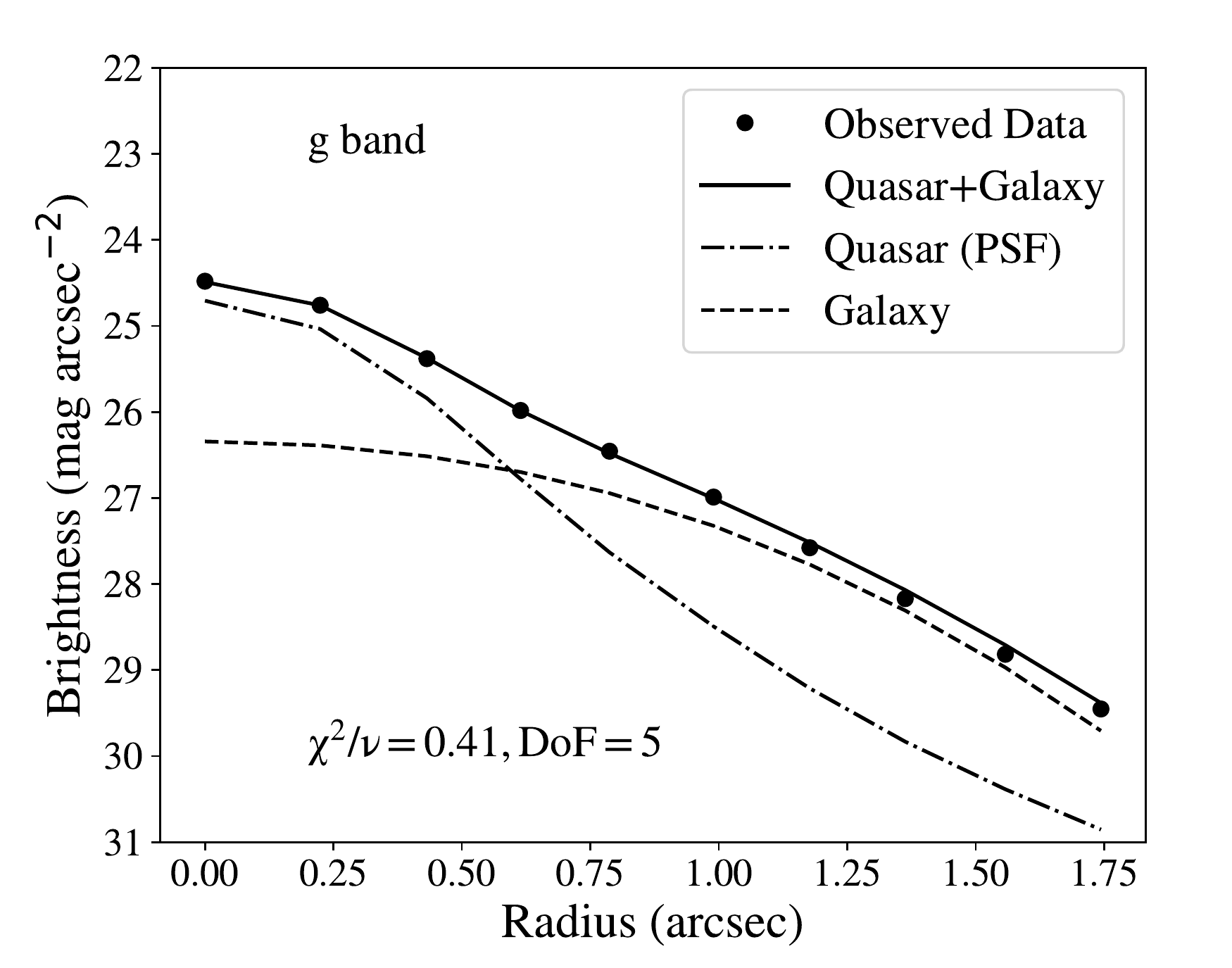}
\end{minipage}
\begin{minipage}{0.49\textwidth}
\raggedright
\includegraphics[trim={0.6cm 0 0 0},width=1.0\textwidth]{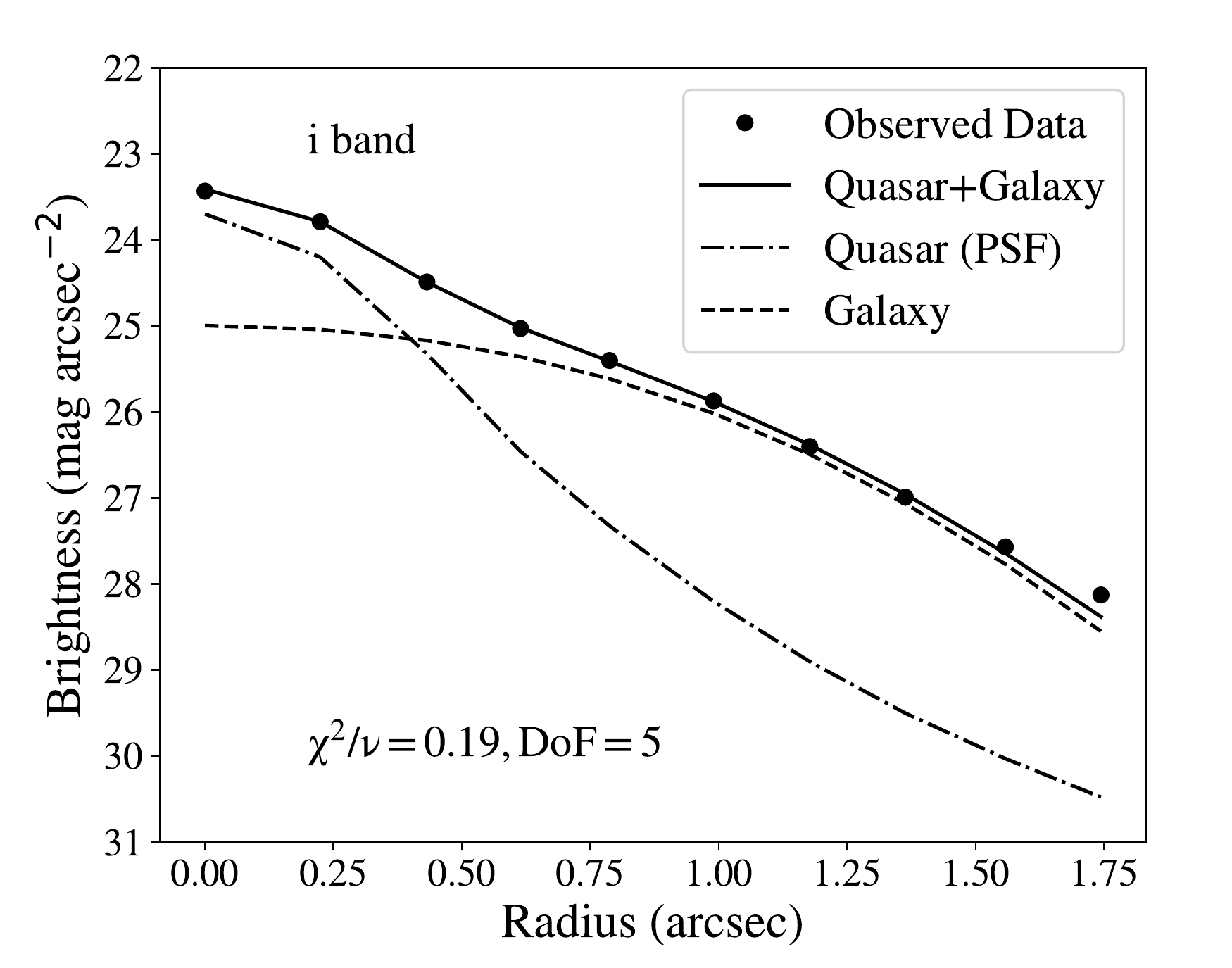}
\end{minipage}
\begin{minipage}{0.49\textwidth}
\raggedleft
\includegraphics[trim={0 0 0.6cm 0},width=1.0\textwidth]{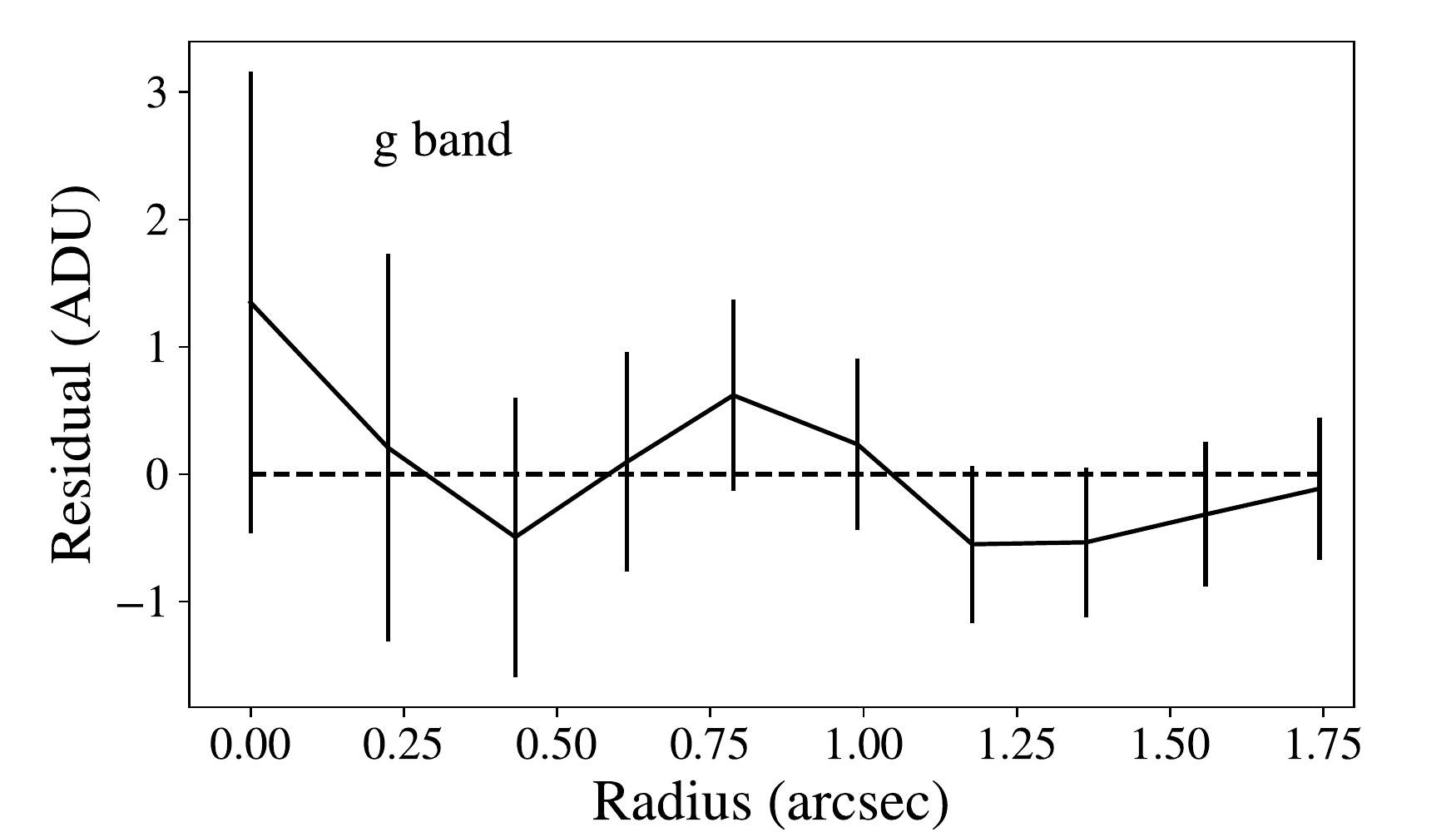}
\end{minipage}
\begin{minipage}{0.49\textwidth}
\raggedright
\includegraphics[trim={0.6cm 0 0 0},width=1.0\textwidth]{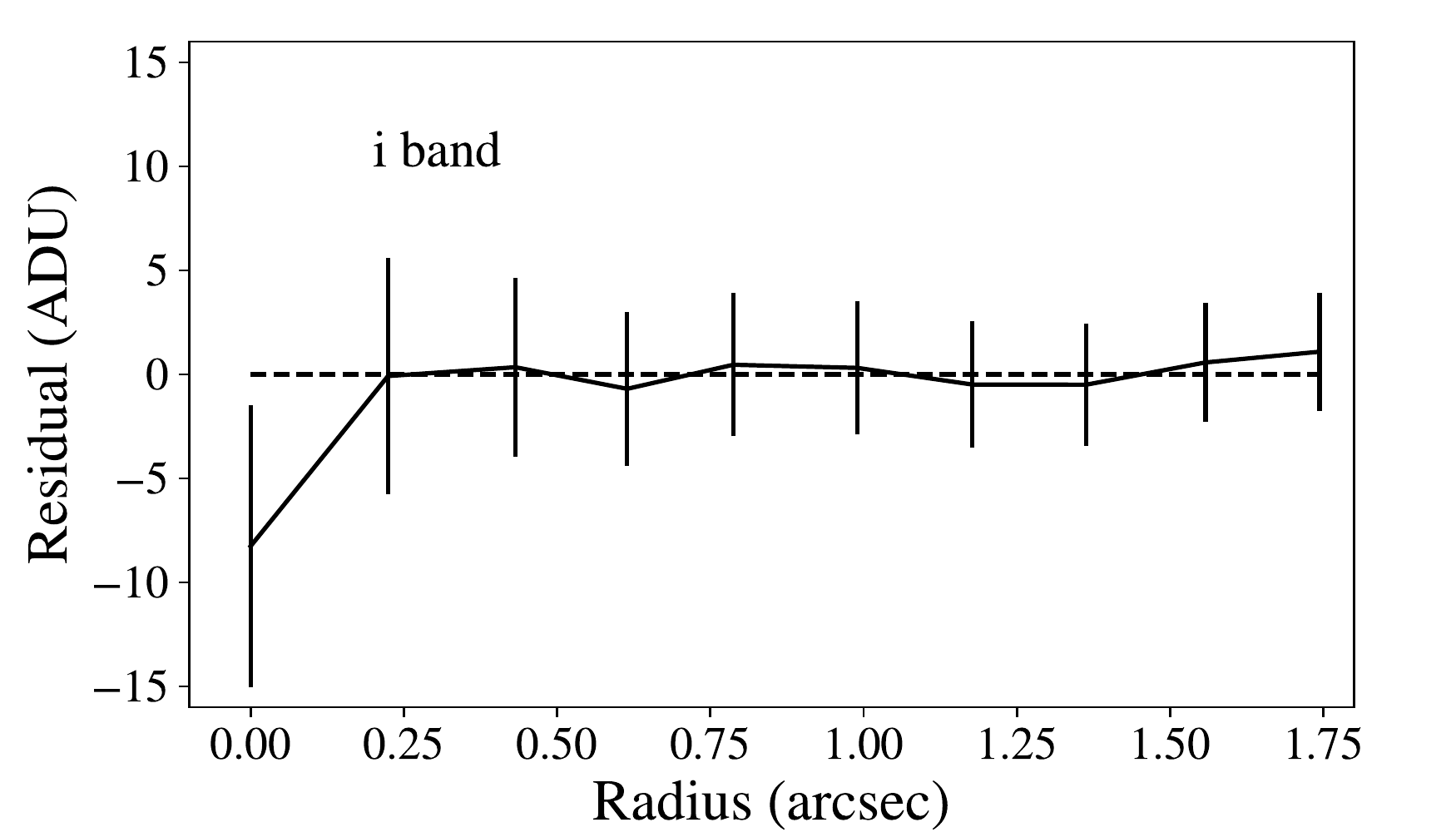}
\end{minipage}
\caption{An example of image decomposition for quasar ID 338 in our sample.
		{\it{Upper Panel}}: the decomposed images. 
		The images are shown in the RGB color mode 
		where the $i$-band images are shown in the R channel, the 
		$g$-band images are shown in the G channel, and the B channel shows nothing. 
		From left to right: the original image, the PSF component, and the PSF-subtracted image.
		The pixel scale of the images is 0.187''/pixel.
		{\it{Middle panels}}: The 1-D image decomposition process in the $g$ and $i$ bands.
		The black dots are the measured 1-D profile of the object.
		The black solid line is the best-fitted 1-D profile.
		The dashed and dot-dashed lines are the best-fitted profiles of the 
		host galaxy and the AGN components, respectively.
		The error bars are very small comparing to the scale of the plot, 
		and they are not shown in this panel for clarity.
		{\it{Lower Panels}}: The residuals of the 1-D profile fitting in the two bands.}
\label{fig:decomp}
\end{figure*}

The fitting procedure involves four parameters: $I_P$, $I_G$, $R_e$, and $n$. 
We allow $R_e$ to vary from 0.5 to 10.0 pixels with a step of 0.1 pixels, and  
$n$ to vary from 0.1 to 5.0 with a step of 0.1. For each pair of $R_e$ and $n$, 
we calculate $I_P$ and $I_G$ using $\partial{\chi^2}/\partial{I_P}=0$ and 
$\partial{\chi^2}/\partial{I_G}=0$ to minimize $\chi^2$. 
For each quasar, we first fit the $i$-band image to obtain the best-fitted $n$ and $R_e$ values. 
We then fix the $n$ and $R_e$ values for the $g$-band image decomposition. 
This is because the $i$-band images have smaller PSF, 
and host galaxies are relatively brighter in the $i$ band. 
Figure \ref{fig:decomp} shows an example of decomposition.
We use the PSF-subtracted images as the best-fitted host galaxy images.
The flux of quasars and galaxies is measured in a $2\arcsec$ aperture,
corresponding to the fiber diameter of the BOSS spectrograph.
The magnitudes, colors, and stellar masses of the
quasar host galaxies that we discuss in Section \ref{discussion} are all based on the $2\arcsec$ 
aperture flux. 
As primary results,
quasar host galaxies in our sample have $g-i$ color $\sim 0.5-2.5$, half-light radius
$R_e \sim 0\farcs 3-1\farcs4$ and S\'ersic index $n\sim 0.5-3$.

\subsection{Comparison to Simulations} \label{simimg}

We assess our decomposition method using simulations.
We use extended objects in our fields, selected from the SDSS photometric catalog,
to mimic galaxy components, and point sources in the fields to mimic AGN components.
The science and noise images of these sources 
are scaled to match the desired galaxy and AGN flux. 
Then these images are combined to make simulated quasar images.
To ensure that the mock ``galaxies" and ``AGNs" can be accurately described by 
S\'ersic profiles and PSF models,
we run the fitting process in Section 3.1 on the mock ``galaxies" and ``AGNs". 
Only ``pure galaxies" with $|m_{\text{S\'ersic}}-m_{\text{galaxy}}|<0.1$ and ``pure AGNs"
with $|m_{\text{PSF}}-m_{\text{AGN}}|<0.1$ are selected for the following analysis.

For the convenience of further discussion, we define several terms and symbols using the $g$
band as an example. 
We use $f^g_\text{gal,\,raw}$ ($f^g_\text{AGN,\,raw}$) to denote the flux of 
extended (point) sources in the $g$ band in the original images,  
and use  $f^g_\text{gal}$ ($f^g_\text{AGN}$) to denote the flux that is scaled to match
the desired galaxy and AGN flux.
We use $R^g_\text{gal}$ to represent the galaxy-to-total flux ratio in the $g$ band, 
$R^g_\text{gal}=f^g_\text{gal}/f^g_\text{total}$, where $f^g_\text{total}=f^g_\text{gal}+f^g_\text{AGN}$. 
We define $f^g_\text{AGN,\,fit}$ as the best-fitted flux of the PSF component. 
The fitting results of galaxies are more complex. 
There are two types of the fitted flux: one is the flux of the model S\'ersic profile, 
which is referred to as $f^g_\text{gal,\,S\'ersic}$.
The other one is the residual flux after the subtraction of the best-fitted PSF component,
which is referred to as $f^g_\text{gal,\,fit}$, i.e., 
$f^g_\text{gal,\,fit}=f^g_\text{total}-f^g_\text{AGN,\,fit}$. 
Accordingly, we define the ``fitted'' galaxy-to-total ratio as $R^g_{\text{gal,\,fit}}=f^g_\text{gal,\,fit}/f^g_\text{total}$.

We first construct a parent sample of simulated host galaxies in the $i$ band.
We generate $10^4$ sets of [$m^i_\text{total}, R^i_\text{gal}$] values 
so that $m^i_\text{total}$ is uniformly distributed between 17 and 23 mag
and $R^i_\text{gal}$ is uniformly distributed between 0 and 1. 
For each pair of [$m^i_\text{total}, R^i_\text{gal}$], 
we calculate $f^i_\text{gal}$ and $f^i_\text{AGN}$, 
and select one extended source and one point source that 
satisfy $|f^i_\text{gal,raw}-f^i_\text{gal}|<0.1\times f^i_\text{gal}$ and 
$|f^i_\text{AGN,raw}-f^i_\text{AGN}|<0.1\times f^i_\text{AGN}$.
Then the two images are scaled so that the flux of the two sources equals
$f^i_\text{gal}$ and $f^i_\text{AGN}$, respectively. 
By doing this, the scaling factors are close to 1, and
the noise of simulated images are close to that of the real data.
The extended sources for the simulated $g$-band images and their scaling factors
are the same as those for the simulated $i$-band images.
The $g$-band flux of the simulated AGN component, $f^g_\text{AGN}$, 
is generated so that the $g-i$ colors of simulated objects 
follow the $g-i$ color distribution of our quasar sample. 
The selection method is the same as for the $i$-band images.
Finally, the images of extended sources and point sources are combined to create the simulated quasar images.


To mimic the real quasar host galaxy sample, 
we select a subset of the parent sample of simulated quasar host galaxies which satisfies 
(1) the distribution of $R_\text{gal}$ is the same as that for the real sample, 
(2) the distribution of the total flux ($f_{\text{total}}=f_{\text{gal}}+f_{\text{AGN}}$), or 
the total magnitude ($m_{\text{total}}$), is the same as the real sample, and 
(3) the distribution of the $g-i$ colors (i.e., $m^g_\text{total}-m^i_\text{total}$) is the same as the real sample. 
The fitting uncertainties are sensitive to the galaxy flux and $R_\text{gal}$.
The simulated sample is used to provide a solid measurement of uncertainties in the fitting process.

We define ``Successful Fitting Criteria" as follows:

(1) Best-fitted S\'ersic index $n>0.1$. 

(2) Best-fitted half-light radius $R_e>1$ pixel.

(3) $(f_\text{gal,\,S\'ersic}-f_\text{gal,\,fit})^2<0.1\times (f_\text{gal,\,fit})^2$ in both $g$ and $i$ bands.

These requirements are set because of the following reasons.
First, an object with $n=0.1$ usually has a very faint galaxy component, 
because the shape of an $n=0.1$ S\'ersic profile is a flat disk at $r<R_e$.
In this case, the best-fitted ``galaxy component'' is likely the residual of background subtraction.
Second, an object with $R_e\le1$ pixel are usually not resolved. Finally,
an object with $(f_\text{gal,\,S\'ersic}-f_\text{gal,\,fit})^2 > 0.1\times (f_\text{gal,\,fit})^2$
(i.e., the flux of the model S\'ersic profile is very different from the total flux minus the PSF component flux)
usually has an unusual morphology which cannot be well described by a S\'ersic profile.
Objects that do not satisfy the criteria 
have large flux fitting error in our simulation
and are rejected in the further analysis.
95 out of 103 quasars in our sample meet the ``Successful Fitting Criteria".

\begin{figure*}
\begin{minipage}{0.5\textwidth}
\raggedleft
\includegraphics[trim={0 0 0.4cm 0},width=0.9\textwidth]{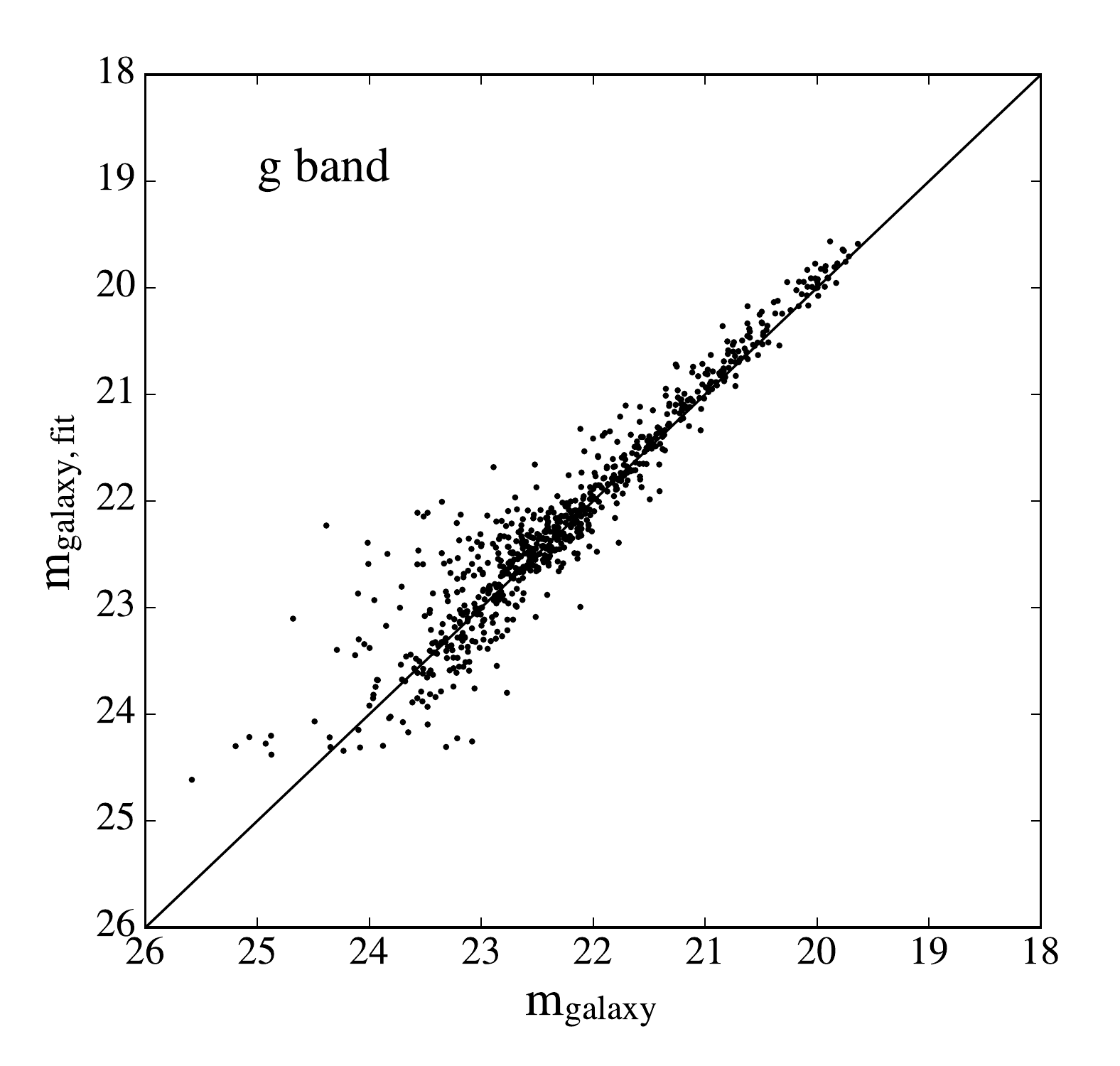}
\end{minipage}
\begin{minipage}{0.5\textwidth}
\raggedright
\includegraphics[trim={0.4cm 0 0 0},width=0.9\textwidth]{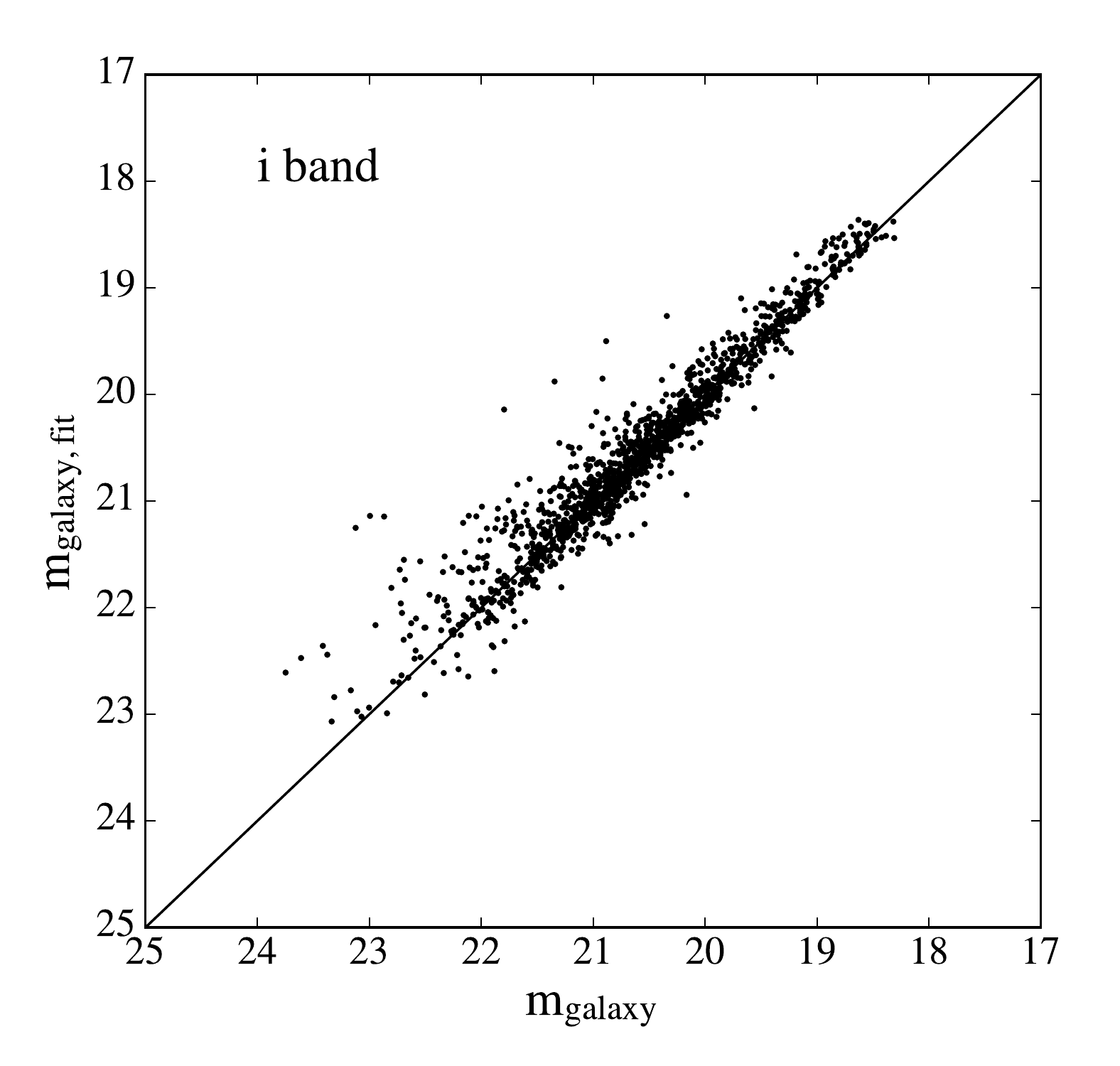}
\end{minipage}
\begin{minipage}{0.5\textwidth}
\raggedleft
\includegraphics[trim={0 0 0 0},width=0.9\textwidth]{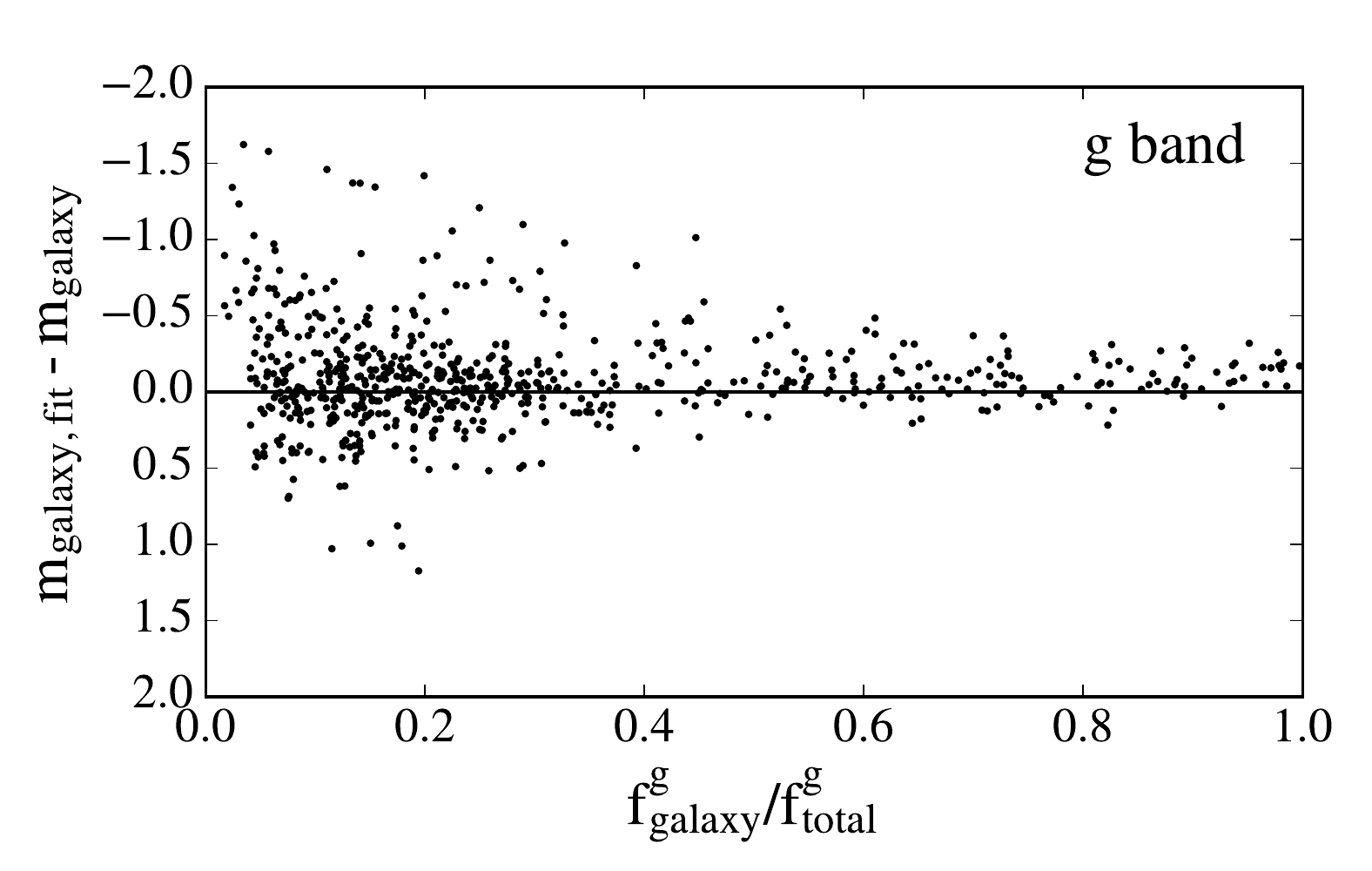}
\end{minipage}
\begin{minipage}{0.5\textwidth}
\raggedright
\includegraphics[trim={0 0 0 0},width=0.9\textwidth]{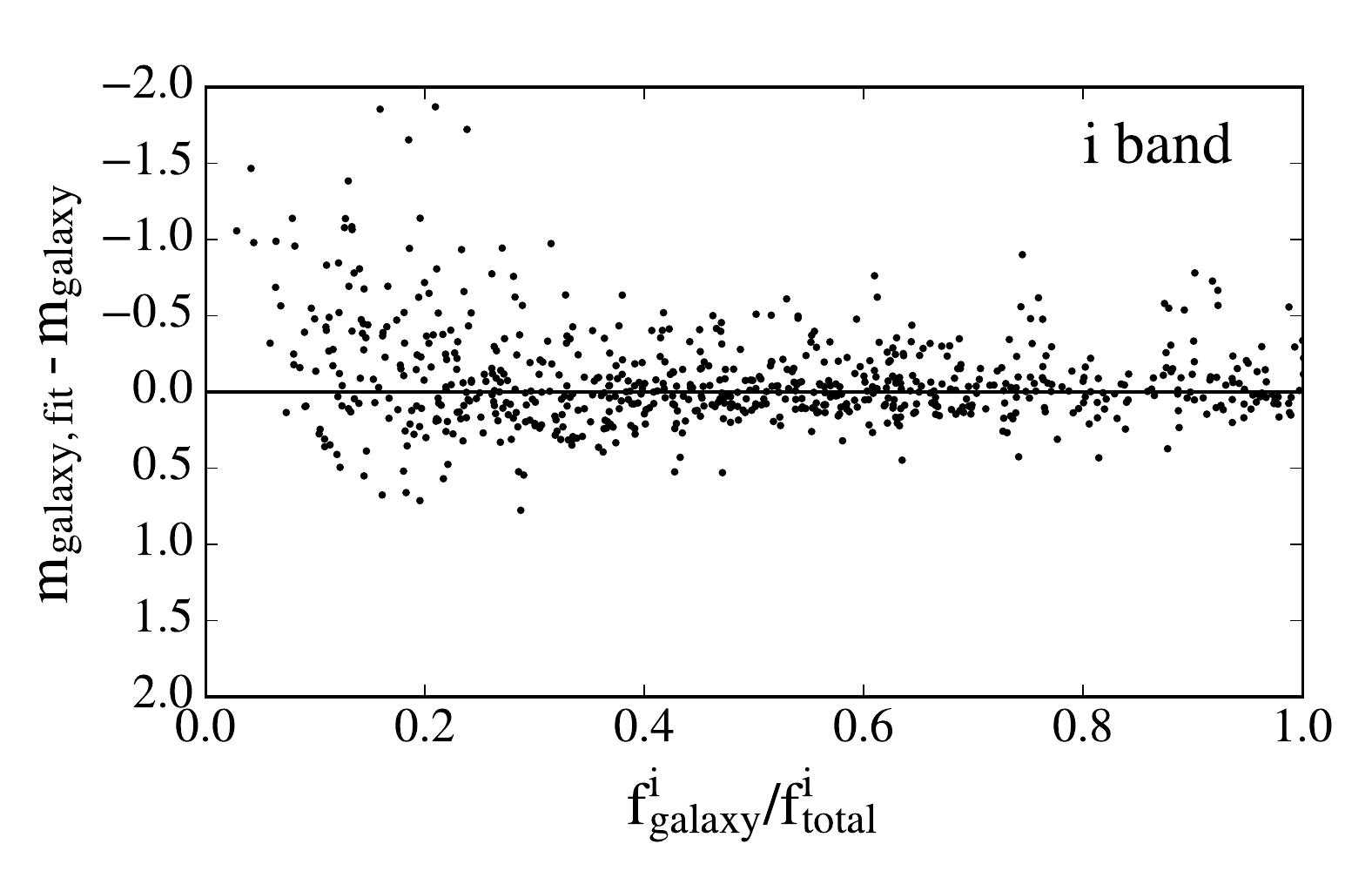}
\end{minipage}
\caption{Estimating the fitting errors of the galaxy magnitudes.
		$Upper$ $Panel:$ Comparison between the real magnitudes and fitted magnitudes of the 
		galaxy components in the simulated quasar host galaxy sample.
		Our fitting technique produces $\Delta m^g_{\text{galaxy}}=-0.04 \pm 0.23$
		and $\Delta m^i_{\text{galaxy}}=-0.02 \pm 0.18$.
		$Lower$ $Panel:$ The influence of the galaxy-to-total flux ratios on the fitting errors.
		At $f^i_{\text{galaxy}}/f^i_{\text{total}}>0.1$,
		there is no evidence that the fitting error evolves with the galaxy-to-total flux ratio,
		while for objects with $f^i_{\text{galaxy}}/f^i_{\text{total}}<0.1$,
		the galaxy flux tends to be slightly overestimated.
		}
\label{fig:magfit}
\end{figure*}

\begin{figure*}
\epsscale{1.1}
\plotone{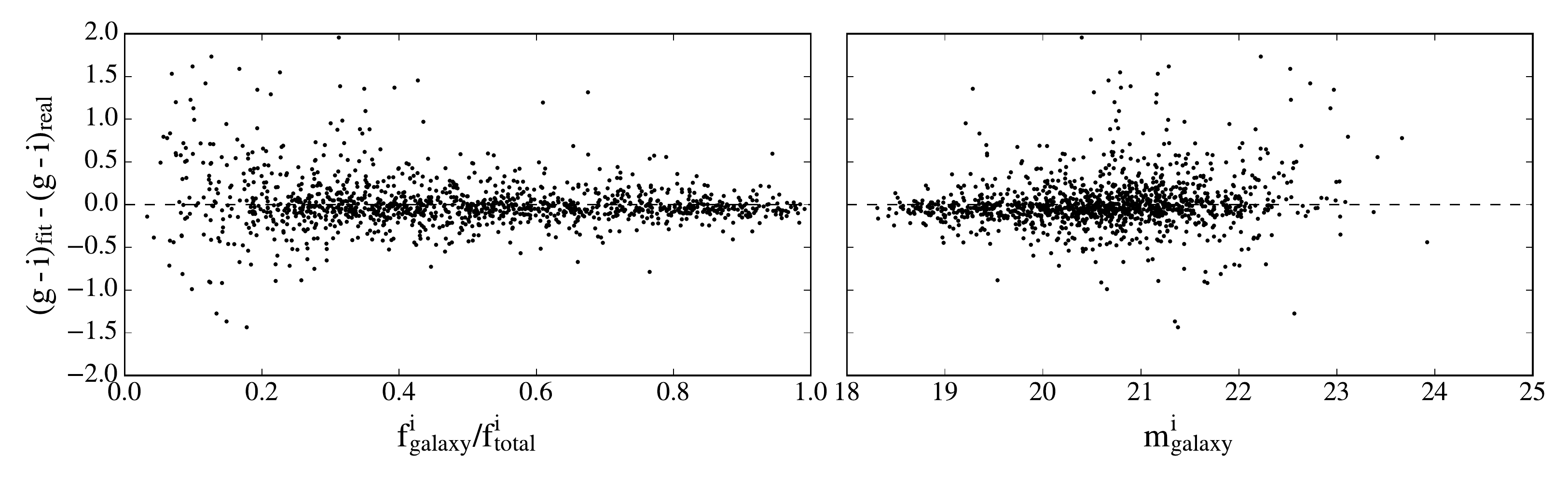}
\caption{Fitting errors of the $g-i$ color of the galaxies.
		$Left$ $Panel:$ The relation between $f^i_{\text{galaxy}}/f^i_{\text{total}}$ and host galaxy $g-i$ color error. 
		Our fitting technique yields $\Delta (g-i)_\text{galaxy}=-0.03 \pm 0.20$. 
		$Right$ $Panel:$ The relation between $m^i_{\text{galaxy}}$ and host galaxy $g-i$ color error. 
		Neither galaxy flux nor galaxy-to-total flux ratio shows obvious influence 
		on the systematic errors of the galaxy color measurement.
		}
\label{fig:g-i}
\end{figure*}

This work focuses on the luminosities, colors, and morphologies of the quasar host galaxies.
We estimate the fitting errors of galaxy magnitudes, colors, half-light radii and S\'ersic indices here.
Since the distributions of the errors are not Gaussian,
we use the so-called ``robust statistical estimators'', i.e.,
the biweight location and the biweight scale \citep{beer90}.
In short, the biweight location and the biweight scale
are counterparts of mean and standard deviation but are less 
sensitive to outliers.
In the following text, the expression $\Delta A=\overline{\Delta A} \pm \sigma_A$ means that the fitting error $\Delta A$
of a quantity $A$ has a biweight location of $\overline{\Delta A}$ and a biweight scale of $\sigma_A$. 
For the successfully fitted objects, the simulation produces $\Delta m_i=-0.02 \pm 0.18$ and $\Delta m_g=-0.04 \pm 0.23$.
The systematic flux errors are much smaller than the random errors.
Figure \ref{fig:magfit} shows the comparison between the real and fitted magnitudes of galaxies 
in the $g$ and $i$ bands. It demonstrates that the systematic fitting errors evolve little with 
the galaxy-to-total flux ratios $R_\text{gal} \equiv f_\text{gal}/f_\text{total}$. 
For objects with $R^g_\text{gal}< 0.05$ or $R^i_\text{gal}< 0.1$, 
our image decomposition tends to overestimate the galaxy flux by $\sim 1$mag.
A similar trend was also reported in \citet{mat14}. 
Therefore, the estimated flux at $R^i_\text{gal,fit} \lesssim 0.2$ 
may suffer larger systematic flux errors,
comparing to the rest of the sample.
There are 8 out of 103 quasars which have $R^i_\text{gal,fit}<0.2$.
We will estimate the typical flux errors of $R^i_\text{gal,fit}<0.2$
quasars and $R^i_\text{gal,fit}>0.2$ quasars respectively in Section \ref{sec:specerr}.

Figure \ref{fig:g-i} presents the systematic errors of the host galaxy $g-i$ colors 
and their dependence on galaxy flux and $R_\text{gal}$.
Our sample gives $(g-i)_{\text{gal,\,fit}}-(g-i)_{\text{gal,\,real}}=-0.03 \pm 0.20$. 
Either galaxy flux or $R_\text{gal}$ has no obvious systematic impact on the measured galaxy colors.

\begin{figure*}[!t]
\epsscale{1.15}
\plotone{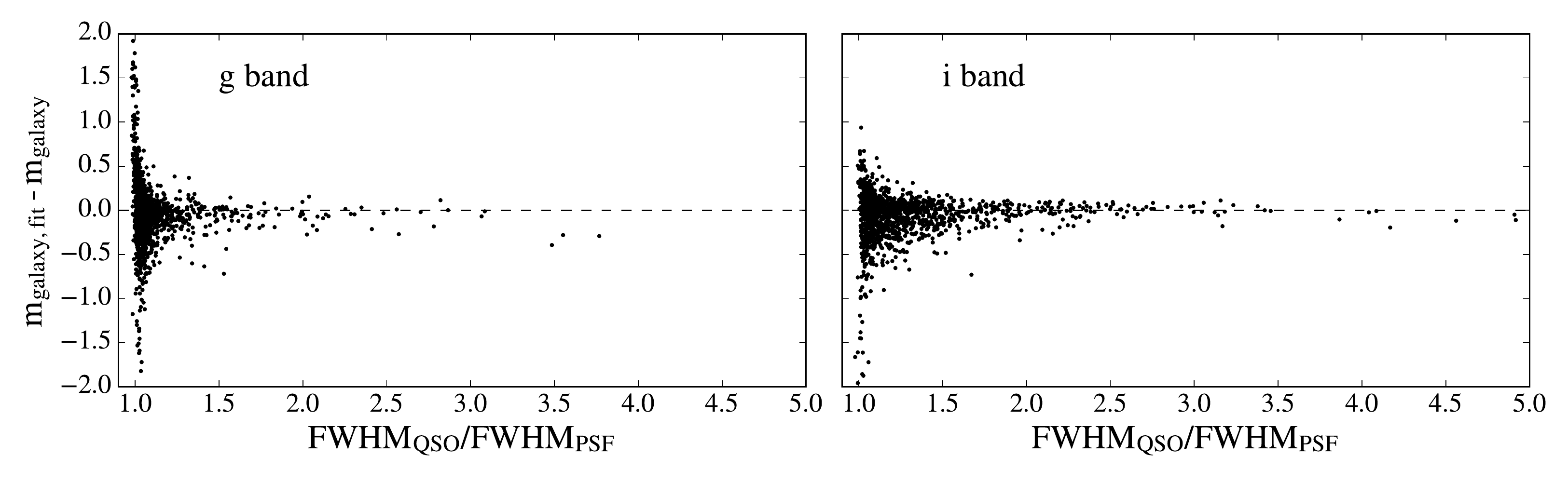}
\caption{The influence of the $\text{FWHM}_\text{QSO}/\text{FWHM}_\text{PSF}$ on the fitting error of the galaxy flux. The flux errors increase towards small $\text{FWHM}_\text{QSO}/\text{FWHM}_\text{PSF}$.
			Outliers with large fitting flux error appear at $\text{FWHM}_\text{QSO}/\text{FWHM}_\text{PSF}<1.05$
			for both bands. 
		}
\label{fig:redFWHM}
\end{figure*}

Figure \ref{fig:redFWHM} illustrates the influence of the quasar-to-PSF FWHM ratios
($\text{FWHM}_\text{QSO}/\text{FWHM}_\text{PSF}$)
on the fitting error of the galaxy flux. 
The random errors increase with decreasing 
$\text{FWHM}_\text{QSO}/\text{FWHM}_\text{PSF}$, as expected.
No significant systematic errors can be seen.
Most outliers in the flux error distribution
appear at $\text{FWHM}_\text{QSO}/\text{FWHM}_\text{PSF}<1.05$ in both bands. 
We compare the distribution of $\text{FWHM}_\text{QSO}/\text{FWHM}_\text{PSF}$ of the simulated sample
with that of the real quasar sample. We use the biweight location and
scale to estimate the distribution of $\text{FWHM}_\text{QSO}/\text{FWHM}_\text{PSF}$. 
The object-to-PSF FWHM ratios in the $i$ band are
$1.17 \pm 0.17$ for the real sample and $1.15 \pm 0.17$ for the simulated sample.
In the $g$ band, the two values are $1.05 \pm 0.05$ and $1.05 \pm 0.05$, respectively.
These results indicates that the real and simulated samples have similar object-to-PSF FWHM ratios,
and thus our error estimation is reliable.

\begin{figure*}
\begin{minipage}{0.5\textwidth}
\centering
\includegraphics[trim={4cm 0cm 4cm 0},width=0.5\textwidth]{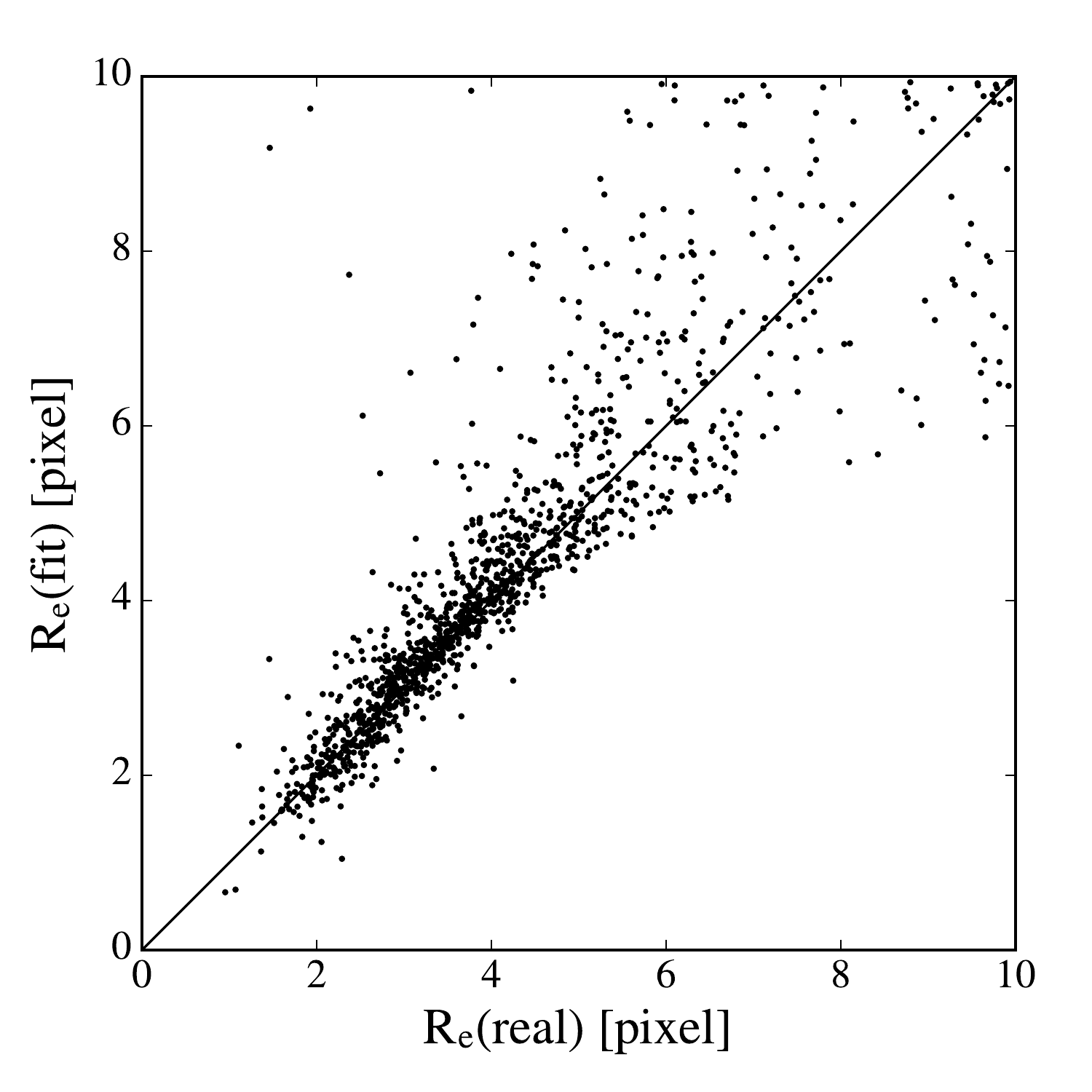}
\end{minipage}
\begin{minipage}{0.5\textwidth}
\centering
\includegraphics[trim={4cm 0cm 4cm 0},width=0.5\textwidth]{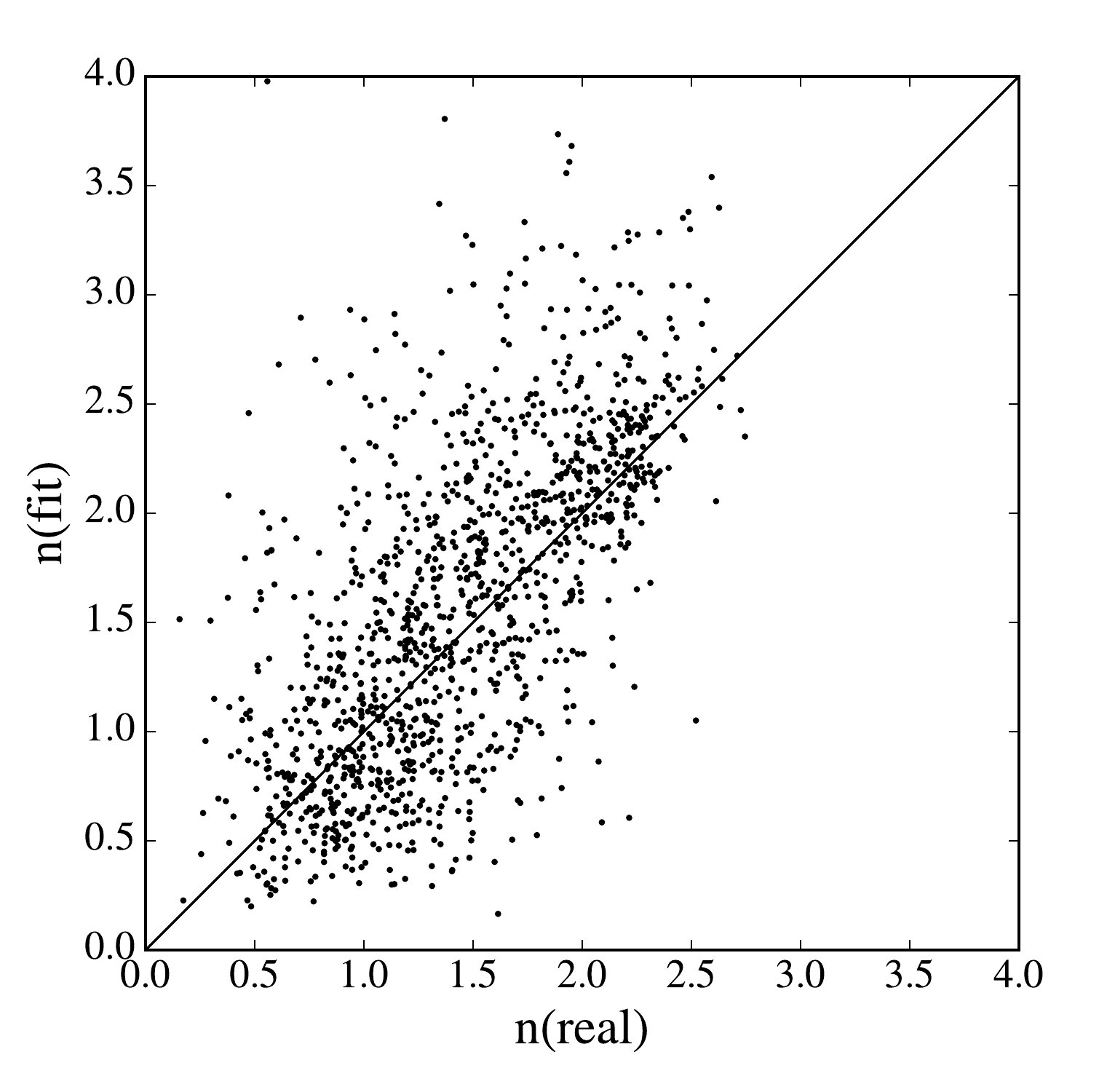}
\end{minipage}
		\caption{Fitting errors of the S\'ersic parameters of galaxies in the $i$ band.
		Random offsets are added to the points to make them distinguishable
		in the grid. $Left$ $Panel:$
        half-light radius $R_e$, with
		 $R_{e,\text{fit}}/R_{e,\text{real}}=1.03 \pm 0.13$.
		$Right$ $Panel:$ 
        S\'ersic index $n$, 
		with $n_{\text{fit}}/n_{\text{real}}=1.06 \pm 0.39$.}
\label{fig:morph}
\end{figure*}

\begin{figure}
\epsscale{1.2}
\plotone{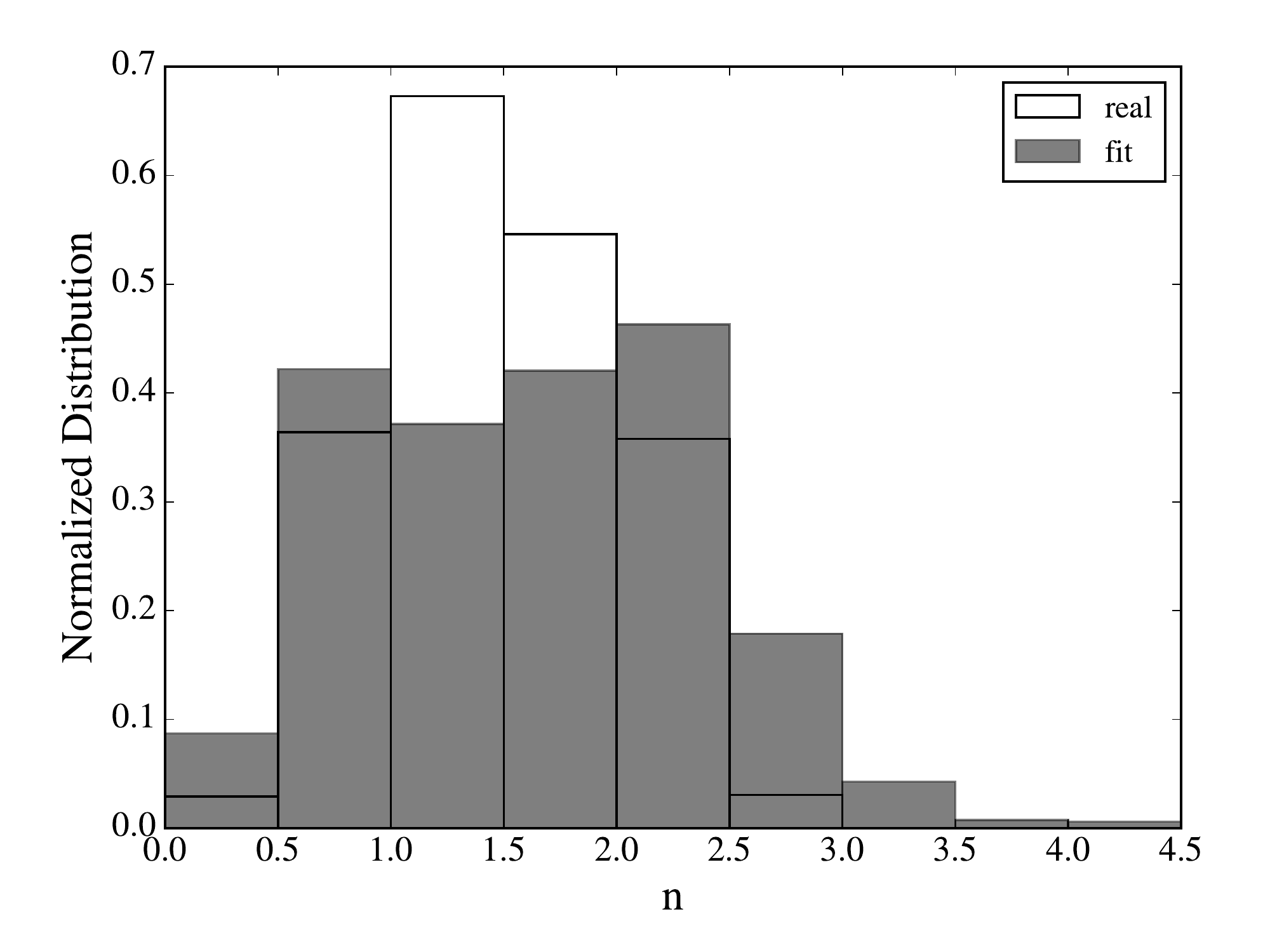}
		\caption{The distribution of ``real" and ``fitted" S\'ersic indices 
		of the simulated quasar host galaxies. The two distributions are roughly consistent.}
\label{fig:nfit}
\end{figure}

Figure \ref{fig:morph} shows the errors of the best-fitted S\'ersic parameters. 
Our decomposition method yields $n_{\text{fit}}/n_{\text{real}}=1.06 \pm 0.39$ 
and $R_{e,\text{fit}}/R_{e,\text{real}}=1.03 \pm 0.13$. 
Although the uncertainties of S\'ersic indices are relatively large, the distribution of $n_{\text{fit}}$
is similar to that of $n_{\text{real}}$ (Figure \ref{fig:nfit}),
which is useful for analyzing the overall quasar host galaxy population.

\subsection{Comparison with Previous Work}
Our image decomposition method is similar to the method used by \citet{mat14}, 
who analyzed the SDSS Stripe 82 images with PSF FWHM $\sim1\farcs1$, and
fitted the 1-D profiles of quasar host galaxies in the five SDSS bands. 
They first fitted the $i$-band images, 
then fitted images in the other bands assuming that the S\'ersic parameters 
($R_e$ and $n$) were the same in all five bands. 
To avoid parameter degeneracy, they fitted the quasar images in two steps.
First, the PSF component was fitted assuming that central pixels 
within $r<2$ pixels are solely produced by the PSF component.
The S\'ersic parameters were fitted after PSF component was subtracted from the original image. 
Their simulations gave $\Delta m_g=-0.08 \pm 0.63$ and $\Delta m_i=0.02 \pm 0.49$. 
There was a clear trend in their simulation that, for host galaxies with small 
$R_\text{gal}$, the fitted galaxy flux was overestimated.
They suggested that their data were not suitable to study host galaxy morphology,
and did not compare the ``real'' and ``fitted'' S\'ersic parameters of their simulated quasar host galaxies.

In our work, we fit the four parameters ($I_P,I_G,R_e,n$) simultaneously.
According to simulation, we set up ``Successful Fitting Criteria'' to exclude objects 
with large fitting error. 
For objects that satisfy the criteria,
the fitting errors are $\Delta m^g_\text{galaxy} = -0.04 \pm 0.23$,
$\Delta m^i_\text{galaxy} = -0.02 \pm 0.18$,
$R_{e,\text{fit}} / R_{e,\text{real}} = 1.03 \pm 0.13$ 
and $n_\text{fit}/n_\text{real} = -0.04 \pm 0.23$.
Our fitting process 
overestimates galaxy flux for objects which have 
$R^g_\text{gal}<0.05$ or $R^i_\text{gal}<0.1$.

\section{Spectroscopic Analysis} \label{spectra}

Traditionally, a $k$-correction is used to convert observed magnitudes to rest-frame magnitudes. 
However, this approach does not work well with only two magnitudes.
We introduce a method that can estimate the rest-frame flux of quasar host galaxies, 
using the results of the image decomposition and 
the high-SNR quasar spectra from the SDSS-RM program.

We use the combined multi-epoch spectra from the SDSS-RM project.
The co-addition strategy can be found in \citet{shen15}.
Briefly, for each quasar, the spectra of 32 epochs were co-added with an
inverse-variance weight. 
The total exposure time of each co-added spectrum is roughly 65 hr.
We then correct for Galactic extinction using the dust map from \citet{Schlegel98} 
and the Galactic extinction curve from \citet{Cardelli89}.
The spectrum of one quasar was severely affected by bad pixels 
and is rejected.

\begin{figure*}
	\epsscale{0.8}
\plotone{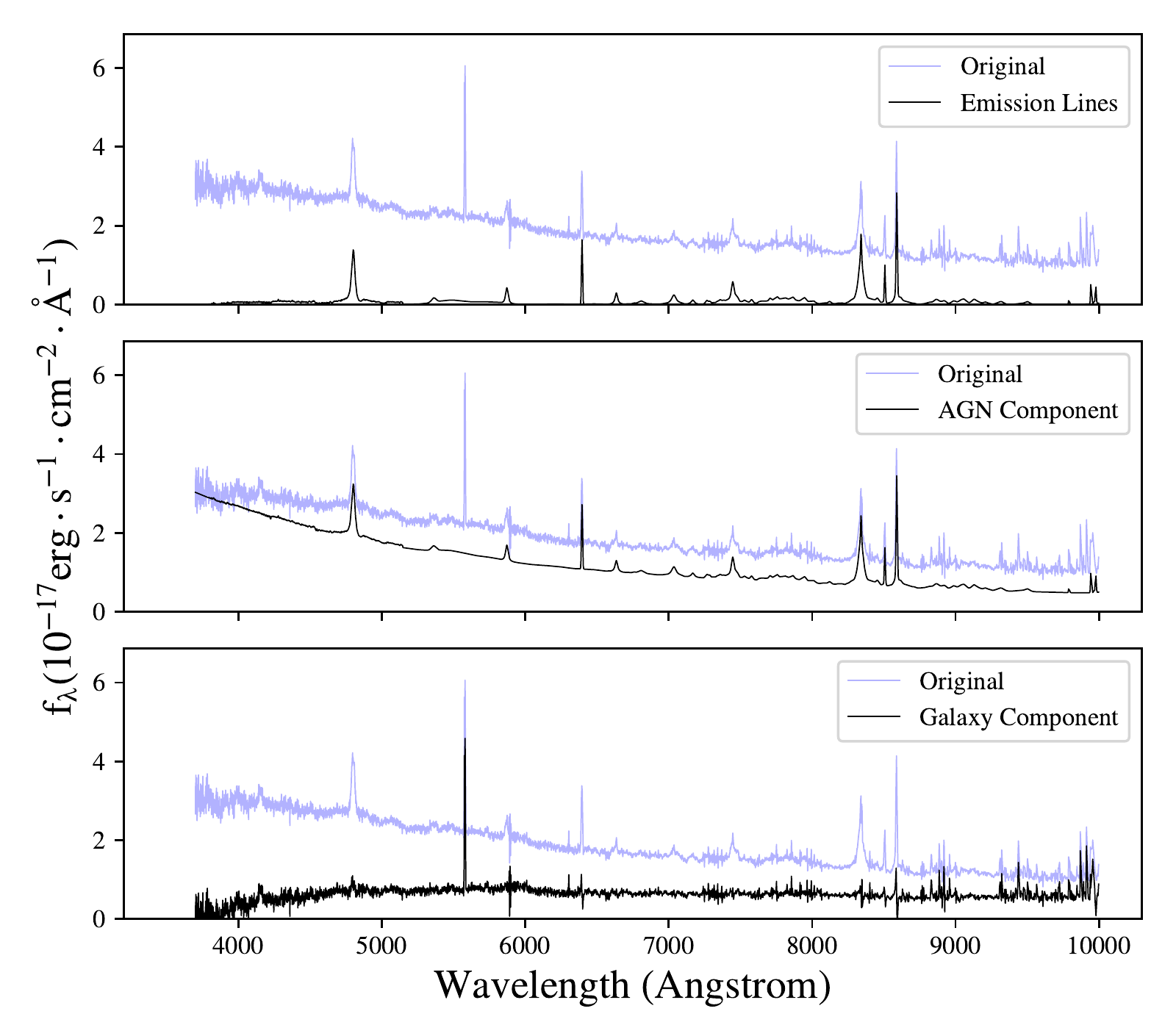}
\caption{An example of quasar host galaxy spectrum fitting ({\texttt{RMID}}=33).
		{\it{Upper Panel}}: Fitting emission lines. 
		The blue line is the original spectrum and the black line is the fitted emission line component. 
		{\it{Middle Panel}}: The modelled AGN component spectrum. 
		Assuming that the AGN spectrum is a power law continuum plus the emission line component, 
		the power law continuum is determined by the AGN-to-total flux ratio in $g$ and $i$ band
		obtained from image decomposition.
		The black line shows the modelled AGN spectrum (power-law continuum plus emission lines).
		{\it{Lower Panel}}: The modelled galaxy spectrum, 
		which is the original spectrum minus the modelled AGN component spectrum.}
\label{fig:conti}
\end{figure*}

\subsection{Method}
The basic idea is to model the AGN component so that 
the AGN-to-total flux ratios in the observed $g$ and $i$ bands 
are equal to the values from the image decomposition.
We assume that the AGN component is described as a power-law continuum plus emission lines,
\begin{equation}
f_\text{AGN}(\lambda)=A\lambda^\alpha+f_\text{lines}(\lambda)
\end{equation}
where the flux of the 
emission lines, $f_\text{lines}(\lambda)$, can be measured by fitting the spectrum.
Under this assumption, the AGN-to-total flux ratio, 
$R^g_\text{AGN}=f^g_\text{AGN}/f^g_\text{total}$, 
is a function of $A$ and $\alpha$.
Solving the equation set
\begin{equation}
\left\{
	\begin{array}{lr}
    R^g_\text{AGN}(A,\alpha) &=R^g_\text{AGN}(\text{image decomposition})\\
    R^i_\text{AGN}(A,\alpha) &=R^i_\text{AGN}(\text{image decomposition})
  	\end{array}
\right.
\end{equation}
gives $A$ and $\alpha$, and thus the AGN spectrum.
The galaxy spectrum is obtained by subtracting the AGN spectrum 
from the total spectrum, and the rest-frame galaxy flux is calculated accordingly.
Given the wavelength range of the SDSS-RM spectra, 
we are able to measure the rest-frame $u$ and $g$ flux for quasars at $0.2<z<0.8$. 

To get the emission line flux,
we fit a rest-frame range 2000 $\sim$ 7200 {\AA} in the spectra of all quasars in our sample. 
The wavelength range covers the observed $g$ and $i$ band at $0.2<z<0.8$. 
We fit nine wavelength intervals separately (see below), 
with each interval fitted as a local power law plus a set of emission lines.
Emission lines (except {\feii} lines) are fitted by Voigt profiles. 
{\feii} lines are modeled by convolving a Gaussian profile with {\feii} templates.
We use {\feii} template from \citet{tsu06} to fit ultraviolet {\feii} lines (2000 $\sim$ 3500 \AA) 
and template from \citet{vc04} to fit optical {\feii} lines (3500 $\sim$ 7000 \AA). 
The fitted emission lines include:

(1) {2000$\sim$3000 \AA: \mgii\ $\lambda2799$ and {\feii} lines.}

(2) {3000$\sim$3500 \AA: {O\,{\sc ii}} $\lambda3134$, He\,{\sc i} $\lambda3188$, 
[Ne\,{\sc v}] $\lambda3347$, 
		[Ne\,{\sc v}] $\lambda3427$ and {\feii} lines.}

(3) {3500$\sim$3900 \AA: [O\,{\sc ii}] $\lambda3726$, [Ne\,{\sc iii}] $\lambda3869$ and {\feii} lines.}

(4) {3900$\sim$4700 \AA: [Ne\,{\sc iii}] $\lambda3967$, H$\delta$, H$\gamma$ and {\feii} lines.}

(5) {4700$\sim$5100 \AA: H$\beta$, {\oiii} $\lambda\lambda4959,5007$ and {\feii} lines.
		One narrow component and one broad component are fitted to the {\hb} emission.}

(6) {5100$\sim$5600 \AA: [Cl\,{\sc iii}] $\lambda5538$ and {\feii} lines.}

(7) {5600$\sim$6200 \AA: He\,{\sc i} $\lambda5876$ and {\feii} lines.}

(8) {6200$\sim$6900 \AA: H$\alpha$, [N\,{\sc ii}] $\lambda6583$, [S\,{\sc ii}] $\lambda\lambda$6716, 6731 lines.
				One narrow component and three broad component are fitted to the {\ha} emission,
				since {\ha} emission features in quasars frequently possess complex line profiles.}

(9) {6900$\sim$7200 \AA: He\,{\sc i} $\lambda7065$. }

Figure \ref{fig:conti} shows an example of emission line fitting and galaxy spectrum modeling.

\begin{figure}
\epsscale{1.2}
\plotone{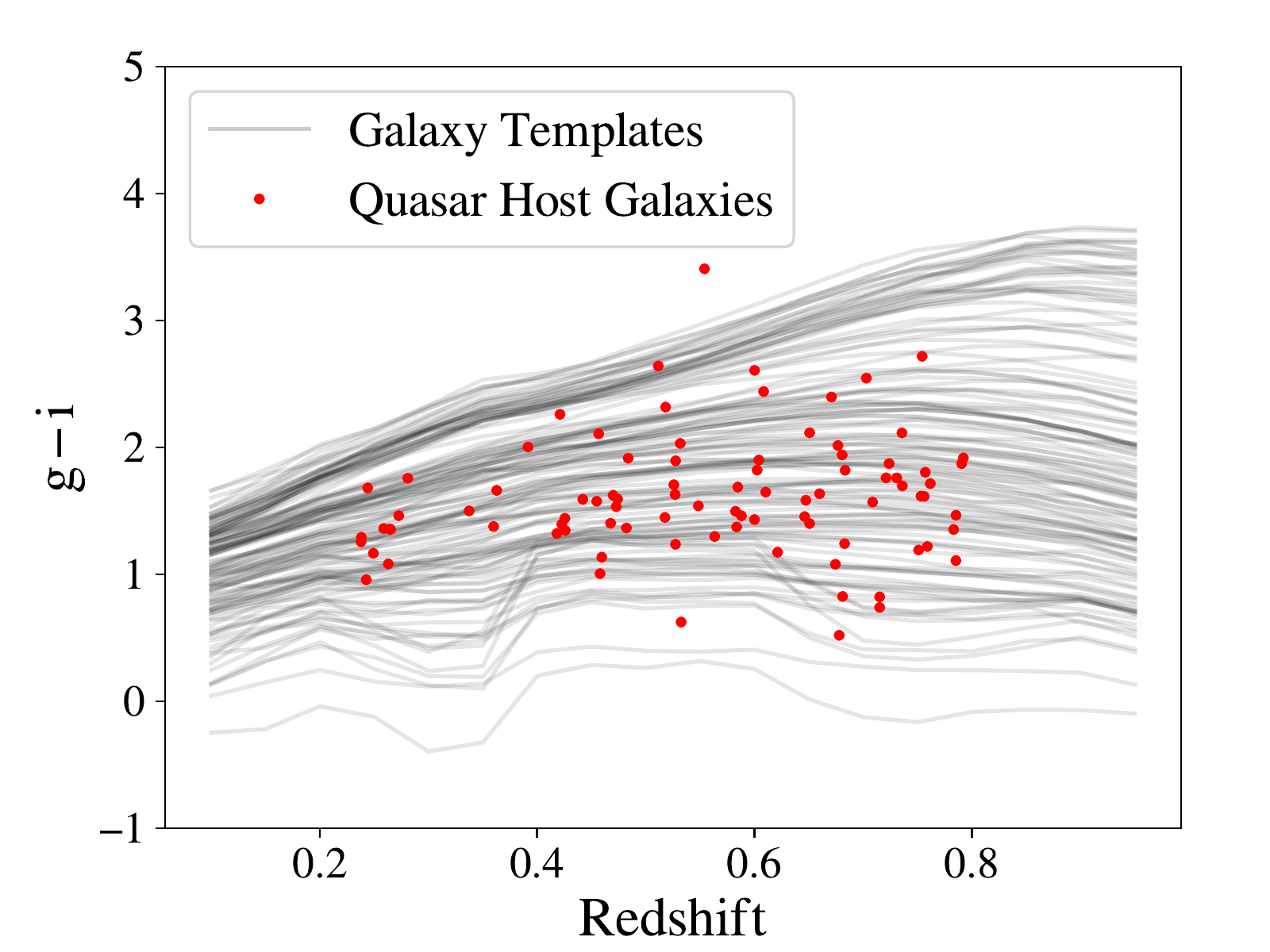}
\caption{The $g-i$ colors of the galaxy templates used to simulate 
the ``AGN + host'' spectra. The grey lines
show the observed $g-i$ colors as a function of redshift. The dots represent the
quasar host galaxies in our sample. The quasar host galaxies and the galaxy templates occupy roughly the
same region in this plot.}
\label{fig:galtemp}
\end{figure}

\begin{figure}
\epsscale{1.2}
\plotone{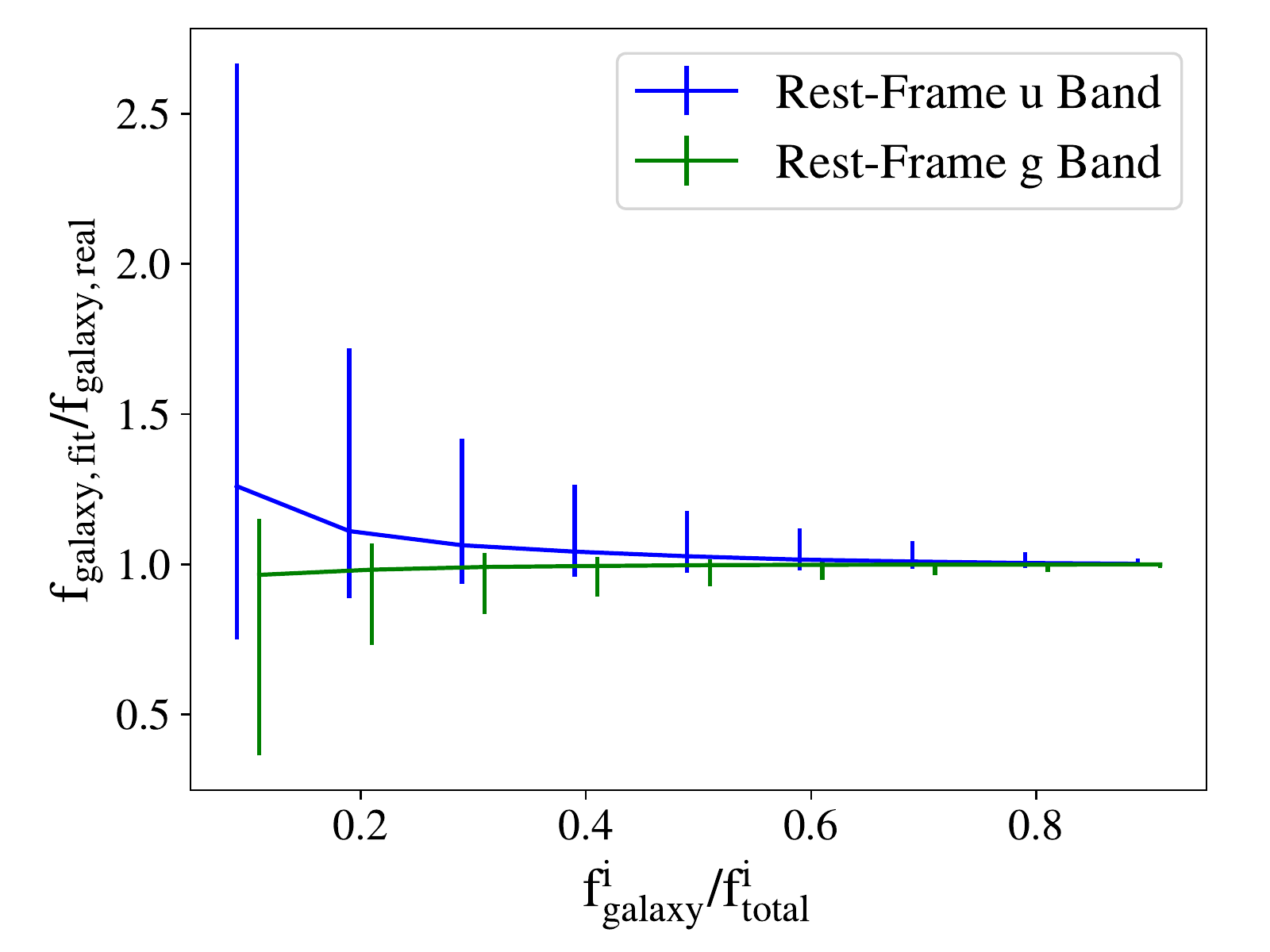}
\caption{The estimated error of galaxy flux from the spectra analysis. The solid lines illustrate the 
		median value and the error bars shows the range that covers 68\% of the simulated spectra. 
		Small offsets are applied on the two lines for the sake of clarity. For objects with 
		$f^i_\text{galaxy}/f^i_\text{total}<0.2$, rest-frame $u$ band flux has a large uncertainty and is likely to be overestimated. 
		The error of the rest-frame $g$-band flux is significantly smaller than that of
		the rest-frame $u$-band flux.
		}
\label{fig:specsim}
\end{figure}

\subsection{Error Estimation} \label{sec:specerr}

We simulate quasar spectra and estimate the errors of the rest-frame galaxy 
flux introduced by our spectroscopic analysis.
We select 12 luminous quasars at $0.2<z<0.8$ from the SDSS DR12 quasar catalog \citep{paris17}
and use their spectra as AGN templates. The details are as follows.
We divide the redshift range $0.2<z<0.8$ evenly into 6 redshift bins, with a bin size of $\Delta z=0.1$, 
and select the brightest two quasars in the $i$ band for each redshift bin.
The two quasars at $0.2<z<0.3$ have $i$-band absolute magnitudes
around --23.5 mag. The quasars in the other redshift bins have $i$-band absolute 
magnitudes brighter than $-24.5$ mag.
We assume that the host galaxy components in these spectra are negligible.
The galaxy templates are from \citet{brown14}, who provided an atlas of high-SNR spectra of 
129 galaxies covering a wavelength range from the rest-frame UV to the mid-IR.
We ensure that the colors of the galaxy templates are close to the colors of the quasar host galaxies
in our sample. Figure \ref{fig:galtemp} shows the observed $g-i$ colors of the quasar 
host galaxies in our sample, compared to the observed $g-i$ colors of the galaxy 
templates. They cover similar parameter space.

We combine the AGN templates and the galaxy templates to generate simulated quasar spectra.
For every possible combination of AGN template and galaxy template, we generate 9 simulated spectra, with
the observed $R^i_\text{gal}$ varies from 0.1 to 0.9 with a step of 0.1. The number of simulated spectra is $12\times129\times9=13932$. 
We apply our spectroscopic analysis to these simulated spectra and calculate the rest-frame $u$ and $g$ flux
of the galaxies.
Figure \ref{fig:specsim} shows the result of the error estimation. 
At $R^i_\text{gal}<0.2$, the rest-frame $u$ band flux is likely to be overestimated, with relatively
large errors. This is mainly due to the difficulty in modeling the small blue bump (SBB) at $\sim 3000${\AA}.
When the galaxy component is very faint compared to the AGN component,
small errors in modeling the SBB will result in large errors in estimating galaxy flux.
At $R^i_\text{gal}>0.2$, the error of the rest-frame $u$-band flux
is comparable to or smaller than the uncertainty from the image decomposition.
The median of $f_\text{gal,\,fit}/f_\text{gal,\,real}$ at $R^i_\text{gal}>0.2$ 
is 1.02 in the rest-frame $u$ band, and the standard deviation is 0.27.
The biweight scale of $f_\text{gal,\,fit}/f_\text{gal,\,real}$ in the rest-frame $u$ band is only 0.006,
meaning that the large error bars shown in Figure \ref{fig:specsim} are mainly from outliers.
The errors in the rest-frame $g$ band are significantly smaller than
those in the rest-frame $u$ band.
At $R^i_\text{gal}<0.2$, the rest-frame $g$ band flux errors 
are comparable to the errors from image decomposition, and the errors decrease towards
larger $R^i_\text{gal}$ values.
At $R^i_\text{gal}>0.2$, the median of $f_\text{gal,\,fit}/f_\text{gal,\,real}$ in 
the rest-frame $g$ band is 0.998, the standard deviation is 0.08, and the biweight scale is 0.0006.
Our simulation shows that,
though the flux errors of  individual quasar host galaxies can be large (especially in the rest-frame $u$ band),
the systematic error are small.

Finally, we estimate typical errors from the combination of our image decomposition and spectroscopic analysis.
As we discussed earlier, objects with $R^i_\text{gal}<0.2$ have significantly larger errors than the rest of the sample, 
so we divide our sample into two subsamples, with $R^i_\text{gal}<0.2$ and $R^i_\text{gal}>0.2$.
The $R^i_\text{gal}<0.2$ subsample consists of only 8 quasars, and 
has large random errors ($\gtrsim 0.5$ mag in both bands).
We focus on their median magnitude and color when interpreting our results, and estimate the 
systematic errors as follows.
According to Figure \ref{fig:magfit}, our image decomposition overestimates 
the flux of these objects by $\sim 0.5$ mag in both observed $g$ and $i$ band, 
and spectroscopic analysis will further overestimate their rest-frame $u$-band flux by $\sim 0.3$ mag.
The net effect is that these objects have their rest-frame $g$-band flux overestimated by $\sim 0.5$ mag,
and their rest-frame $u-g$ colors underestimated by $\sim 0.3$ mag.

The $R^i_\text{gal}>0.2$ subsample does not have significant systematic flux errors.
Our simulated sample shows that
the uncertainties of the observed $g$- and $i$-band magnitudes are 0.21 and 0.17 mag,
respectively.
We use 0.2 mag as the typical error of image decomposition.
The uncertainty of spectroscopic analysis is negligible, or comparable to the uncertainty from
the image decomposition. For the typical error of spectroscopic analysis,
we take the standard deviation as a conservative estimate, which is 0.26 mag for the rest-frame $u$ band
and 0.08 mag for the rest-frame $g$ band.
We then take the quadrature sum of the uncertainties from the two steps as final magnitude uncertainty,
which gives $\sigma_u=0.33$ and $\sigma_g=0.22$ for the two bands.
The errors of other properties derived from flux, including the rest-frame $u-g$ colors and stellar masses, are estimated accordingly.

\section{Results} \label{results} 

\begin{figure}
\epsscale{1.2}
\plotone{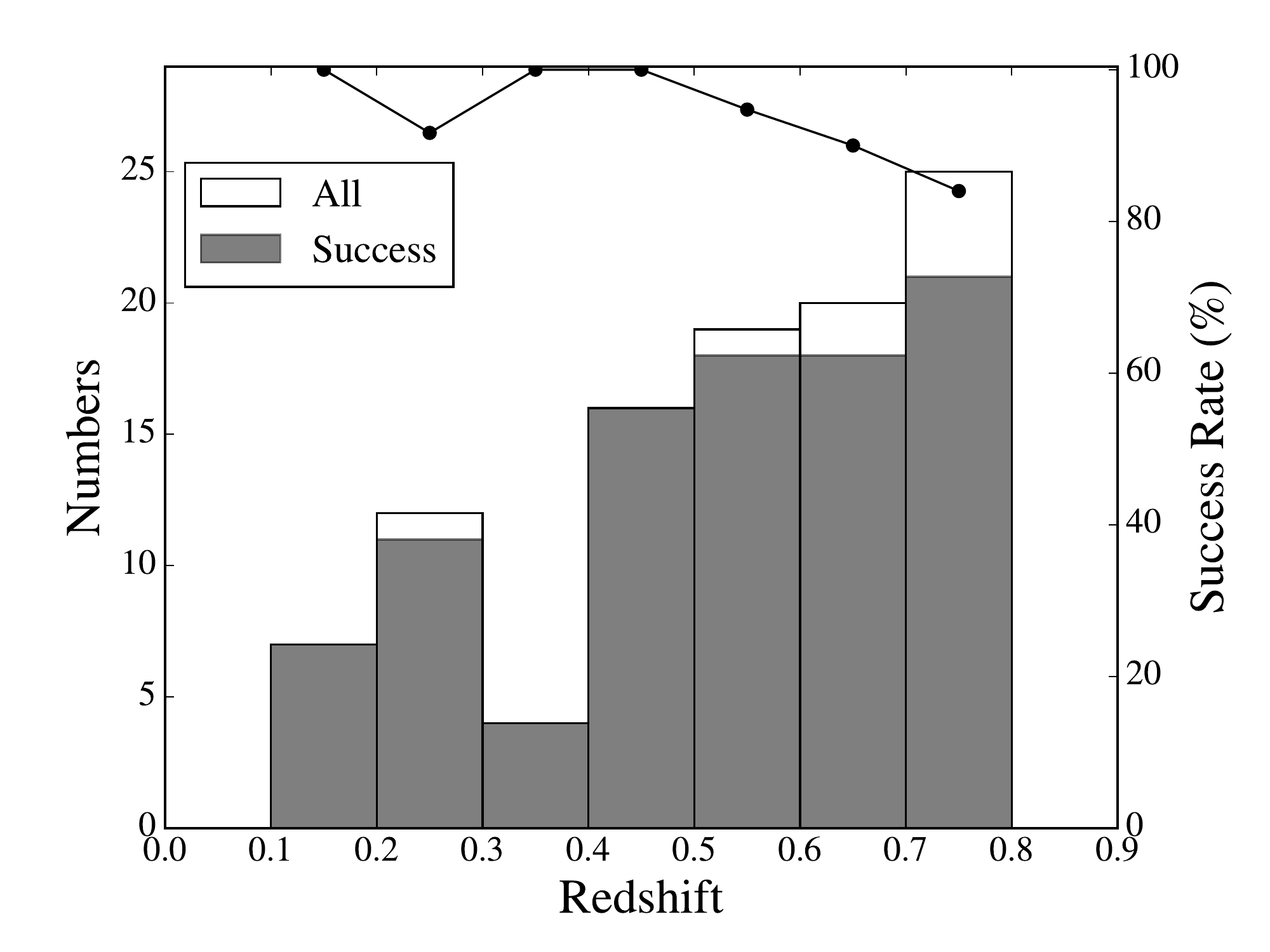}
\caption{The ``success rate" of the image decomposition as a function of redshift. 
		The histogram displays the total number (white) and 
		the number of successfully fitted quasars (gray) in each redshift bin. 
		The filled circles indicate the success rates in individual redshift bins. 
		A total of 88 out of 96 quasars in our $0.2<z<0.8$ sample are successfully fitted. 
		The success rate decreases slowly with redshift, 
		and the success rate at $0.7<z<0.8$ is 84\%. }
\label{fig:success}
\end{figure}

We fit all 103 quasars in our sample, and 95 of them meet the 
``Successful Fitting Criteria'' defined in Section \ref{simimg}.
Figure \ref{fig:success} shows the success rate as a function of redshift. 
The success rate decreases slowly with redshift.
The success rate at $0.7<z<0.8$ is 84\%, which is still high.
This indicates that our redshift cut ($z<0.8$) is reasonable.
Our optical spectra do not cover the rest-frame $u$ band for quasars at $z<0.2$,
so we focus on the quasars at $0.2<z<0.8$. 
There are 95 quasars in this redshift range, and 87 of them are successfully fitted.

\subsection{Flux and Colors of Quasar Host Galaxies} \label{fluxcolor}

\begin{figure}
\epsscale{1.2}
\plotone{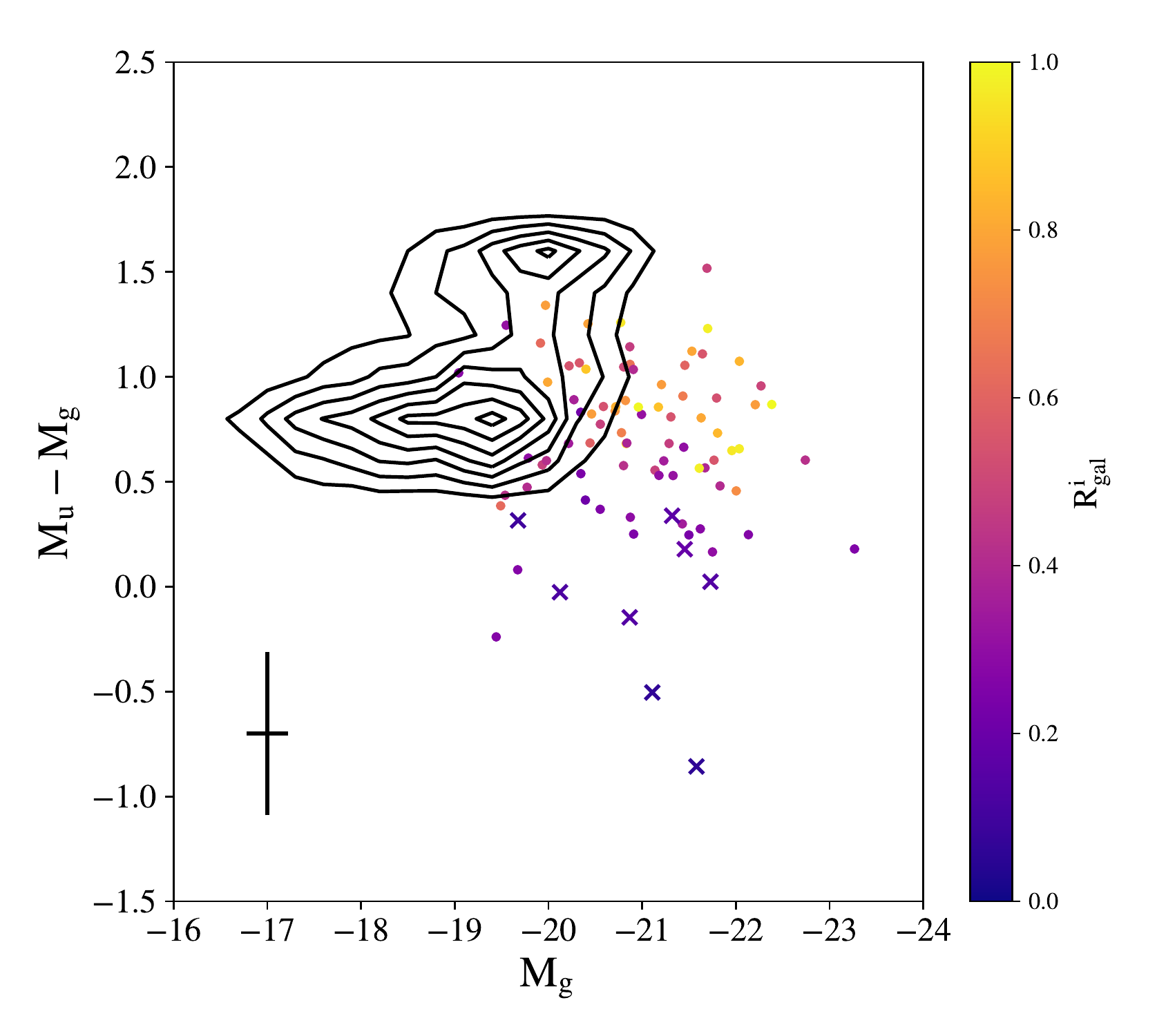}
\caption{The rest-frame CMD ($M_g$ vs $M_u-M_g$) of the quasar host galaxies. 
		The crosses represent the quasar host galaxies with $R^i_\text{gal}<0.2$,
		and the black dots represent the host galaxies with $R^i_\text{gal}>0.2$.
		The contour shows the galaxies at $0.2<z<0.8$ from the 
		COSMOS/UltraVISTA $K$-selected galaxy catalog. 
		Compared to these normal galaxies, our
		quasar host galaxies are more luminous.
		The $u-g$ colors of the host galaxies are similar to those of 
		star-forming galaxies.
		}
\label{fig:mgug}
\end{figure}

\begin{figure}
\epsscale{1.2}
\plotone{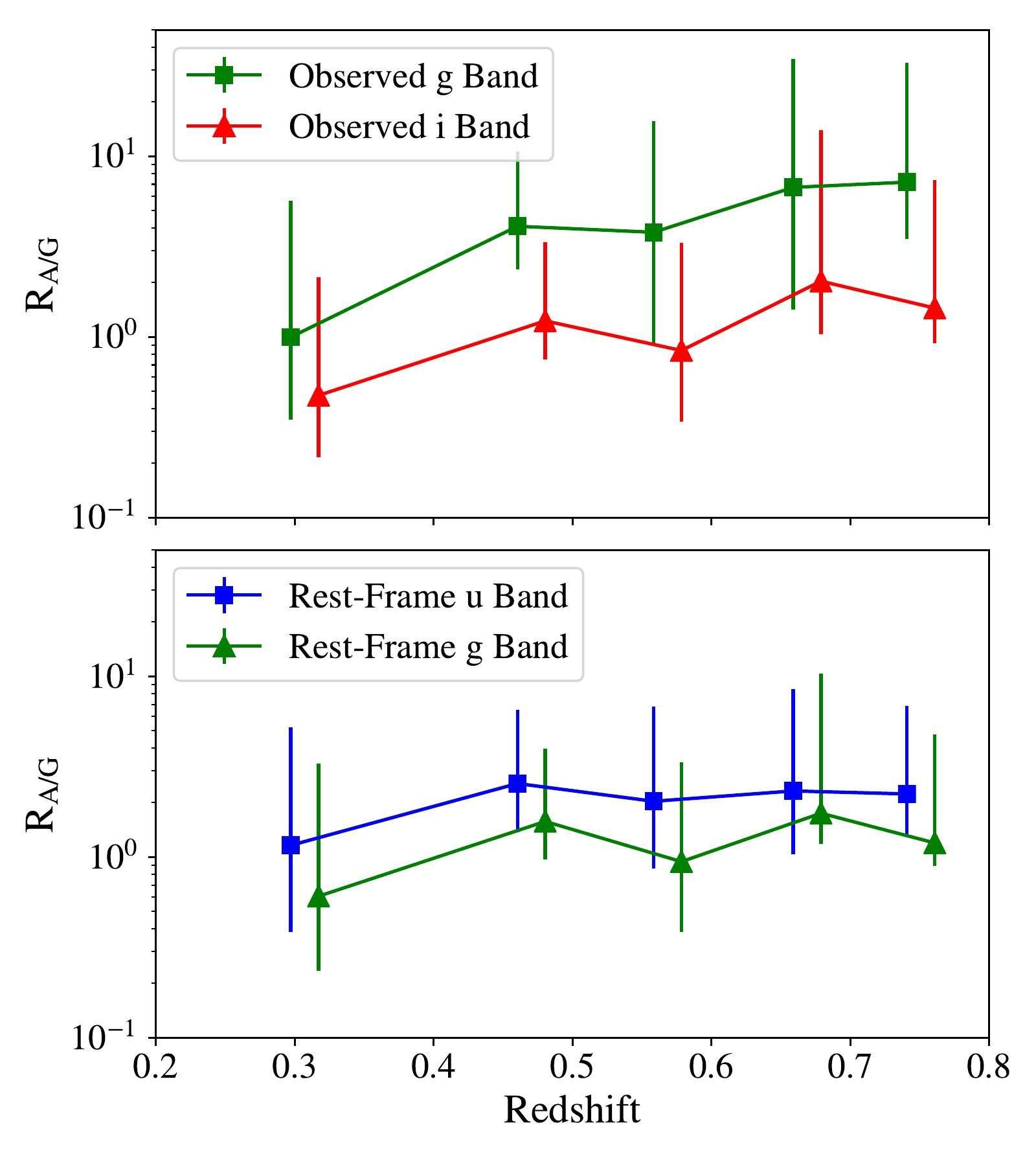}
\caption{The AGN-to-galaxy flux ratio ($R_\text{A/G}$).
		Each redshift bin has the same number of quasars.
		The points and the error bars show the median values and the ranges where $68\%$
		of the objects are included.
		The redshifts are shifted slightly to make the error bars of different lines distinguishable.
		{\it{Upper Panel}}: $R_\text{A/G}$ in the $observed$ $g$ and $i$ bands. 
		At $0.2<z<0.8$, $R_\text{A/G}$ increases with redshift. 
		{\it{Lower Panel}}: $R_\text{A/G}$ in the {\it{rest-frame}} $u$ and $g$ bands. 
		They show little evolution with redshift.}
\label{fig:ratio}
\end{figure}

We calculate the rest-frame $u$- and $g$-band absolute magnitudes of the quasar host galaxies
based on the host galaxy spectra obtained in our spectroscopic analysis.
The median $u-g$ color of the sample is $0.68$ and the standard deviation is $0.40$.
Figure \ref{fig:mgug} shows the color-magnitude diagram (CMD) of these host galaxies. 
The crosses represent the 8 galaxies with $R^i_\text{gal}<0.2$, and the dots represent the 
galaxies with $R^i_\text{gal}>0.2$. 
In this figure we also plot the distribution of $0.2<z<0.8$ galaxies 
from the COSMOS/UltraVISTA $K$-selected galaxy catalog \citep{muzzin13}. 
Compared to these normal galaxies, 
our quasar host galaxies occupy a different region in the CMD:
the host galaxies are significantly more luminous. 
On the other hand, their global $u-g$ colors are similar to those of star forming galaxies 
(i.e., galaxies located in the ``blue cloud''). 
All the above suggests that the quasar host galaxies  
in our sample are mostly luminous star-forming galaxies, which
is in consistent with positive AGN feedback.
The blue end of the $u-g$ colors are dominated by several objects at $R^i_\text{gal}<0.2$.
As we discussed earlier, these extreme colors are likely caused by the bias and 
large uncertainties from the imaging decomposition and spectroscopic analysis.
Besides those with $R^i_\text{gal}<0.2$, there are still some objects
which are extremely blue $(u-g\sim -0.2)$, yet they are consistent with the ``blue cloud''
in $\sim2\sigma$ level.

Figure \ref{fig:ratio} shows the AGN-to-galaxy flux ratio 
($R_\text{A/G}$) as the function of redshift. 
We divide the quasar sample into five redshift bins, 
with each bin having the same number of quasars.
From $z \sim 0.3$ to $\sim 0.7$, the median $R_\text{A/G}$ values increase from $\sim 1$ to $\sim 7$ in the observed $g$ band, and from $\sim 0.5$ to $\sim 1$ in the observed $i$ band. 
The lower panel of the figure shows $R_\text{A/G}$ in the rest-frame $u$ and $g$ bands.
In the $u$ band, $R_\text{A/G}=2.2^{+4.7}_{-0.8}$, and $R_\text{A/G}=1.2^{+3.2}_{-0.3}$
in the $g$ band. As expected, $R_\text{A/G}$ is larger in the $u$ band. The $R_\text{A/G}$ values
also suggest that host galaxies are significant in these quasars.
There is no obvious trend of $R_\text{A/G}$ with redshift.
The redshift-dependence in the observed $g$ and $i$ bands 
mainly arises from the fact that the two observed bands cover different 
rest-frame wavelength ranges for quasars at different redshifts. 

\subsection{$M_{*}-M_\text{BH}$ Relation} \label{BHstar}

We calculate the stellar masses of the quasar host galaxies 
using the stellar mass-to-light ratio from \citet{bell03}:
\begin{equation}
\label{eq:mstar}
\text{log}(M_*/L_g)=-0.221+0.485 \times (u-g),
\end{equation}
where $M_*$ and $L_g$ are in the solar units. 

The black hole mass of the quasars are adopted from Shen et al. (in preparation),
who use the luminosity at 5100{\AA} $(L_{\text{5100}})$ 
and the line width of {\hb}, based on the empirical relation by \citet{vp06}:
\begin{equation}
\label{eq:mbh}
\begin{split}
\text{log}(\frac{M_\text{BH}}{M_{\odot}})=a+b~\text{log}(\frac{L}{10^{44}\text{ erg s}^{-1}})+c~\text{log}(\frac{\text{FWHM}}{\text{km s}^{-1}}),
\end{split}
\end{equation}
where $a=0.91, b=0.50$, and $c=2$ when using the broad {\hb} line and
the AGN luminosity at 5100{\AA}, $L_{\text{5100}}$.
Shen et al. (in preparation) does not consider the contribution of galaxy fluxes
in $L_{\text{5100}}$, which is corrected in this work according to the 
decomposed spectra in Section \ref{spectra}.

\begin{figure}
\centering
\includegraphics[trim={1cm 0 0.5cm 0},width=0.5\textwidth]{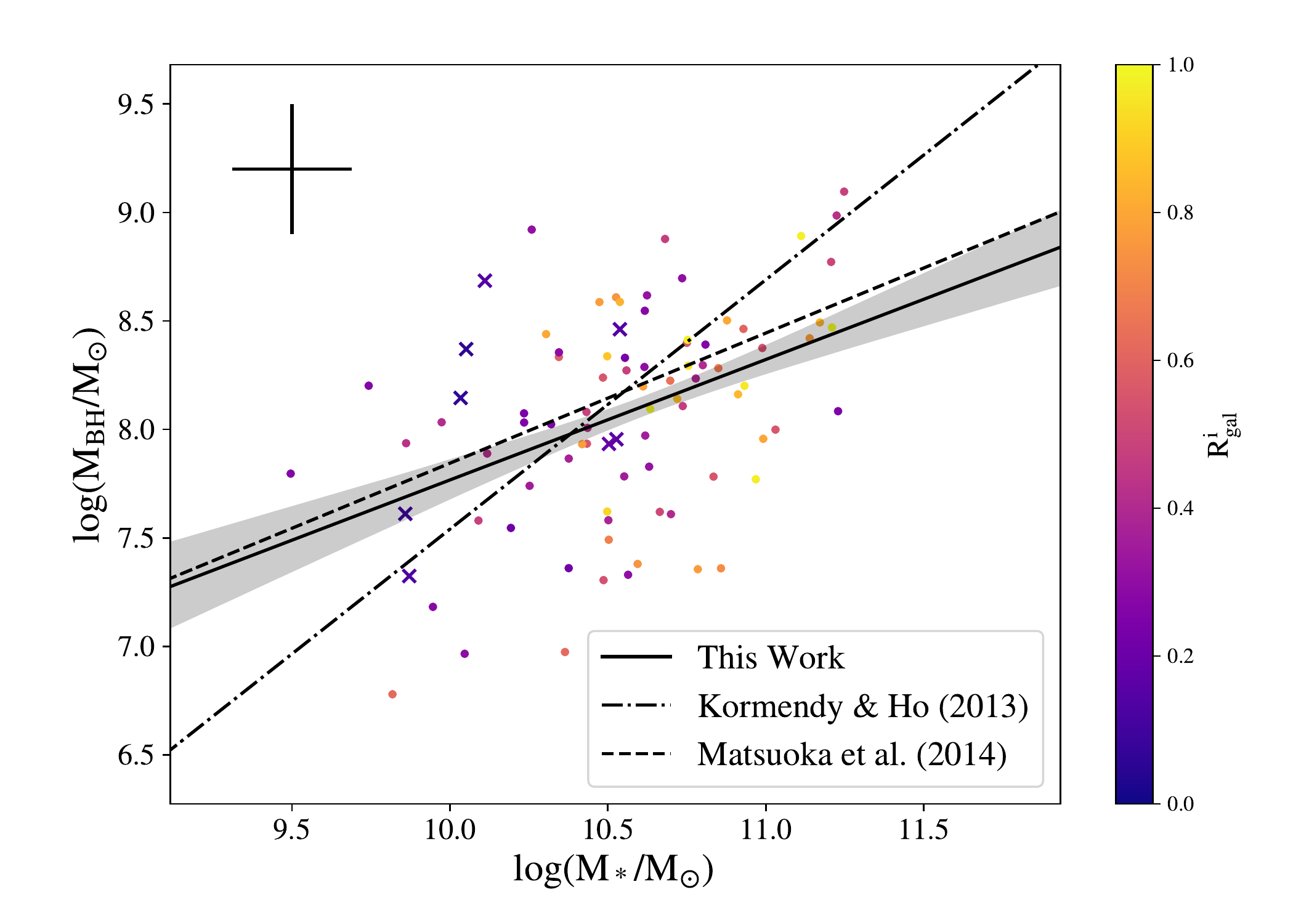}
\caption{The $M_{*} - M_{\text{BH}}$ relation in the quasar host galaxies. 
		The crosses represent the quasars with $R^i_\text{gal}<0.2$,
		and the black dots shows the quasars with $R^i_\text{gal}>0.2$.
		The solid line shows the best fit for all quasars, 
		the dashed line shows the $M_{\text{Bulge}} - M_{\text{BH}}$ relation in local galaxies
		from \citet{kh13}, and the dot-dashed line represents the relation 
		for $z<0.6$ quasar host galaxies from \citet{mat14}. 
        The gray area shows the $1\sigma$ error of our fitting result estimated using Bootstrap.
        This figure suggests that 
the $M_{*} - M_\text{BH}$ of quasar host galaxies are shallower than the
$M_{\text{Bulge}} - M_{\text{BH}}$ relation in local galaxies.
		}
\label{fig:mbh}
\end{figure}

Figure \ref{fig:mbh} shows the $M_{*} - M_{\text{BH}}$ relation of the quasar host galaxies. 
There is a positive correlation between $M_{*}$ and $M_{\text{BH}}$.
We also include the $M_{\text{Bulge}} - M_{\text{BH}}$ relation in local galaxies from \citet{kh13}
and the relation for $z<0.6$ SDSS quasars from \citet{mat14}.
The stellar masses in our sample and in the \citet{mat14} sample 
include both bulge and disk masses. Figure \ref{fig:mbh} suggests that 
the $M_{*} - M_\text{BH}$ of quasar host galaxies are shallower than the
$M_{\text{Bulge}} - M_{\text{BH}}$ relation in local galaxies.

\begin{figure}
\epsscale{1.1}
\plotone{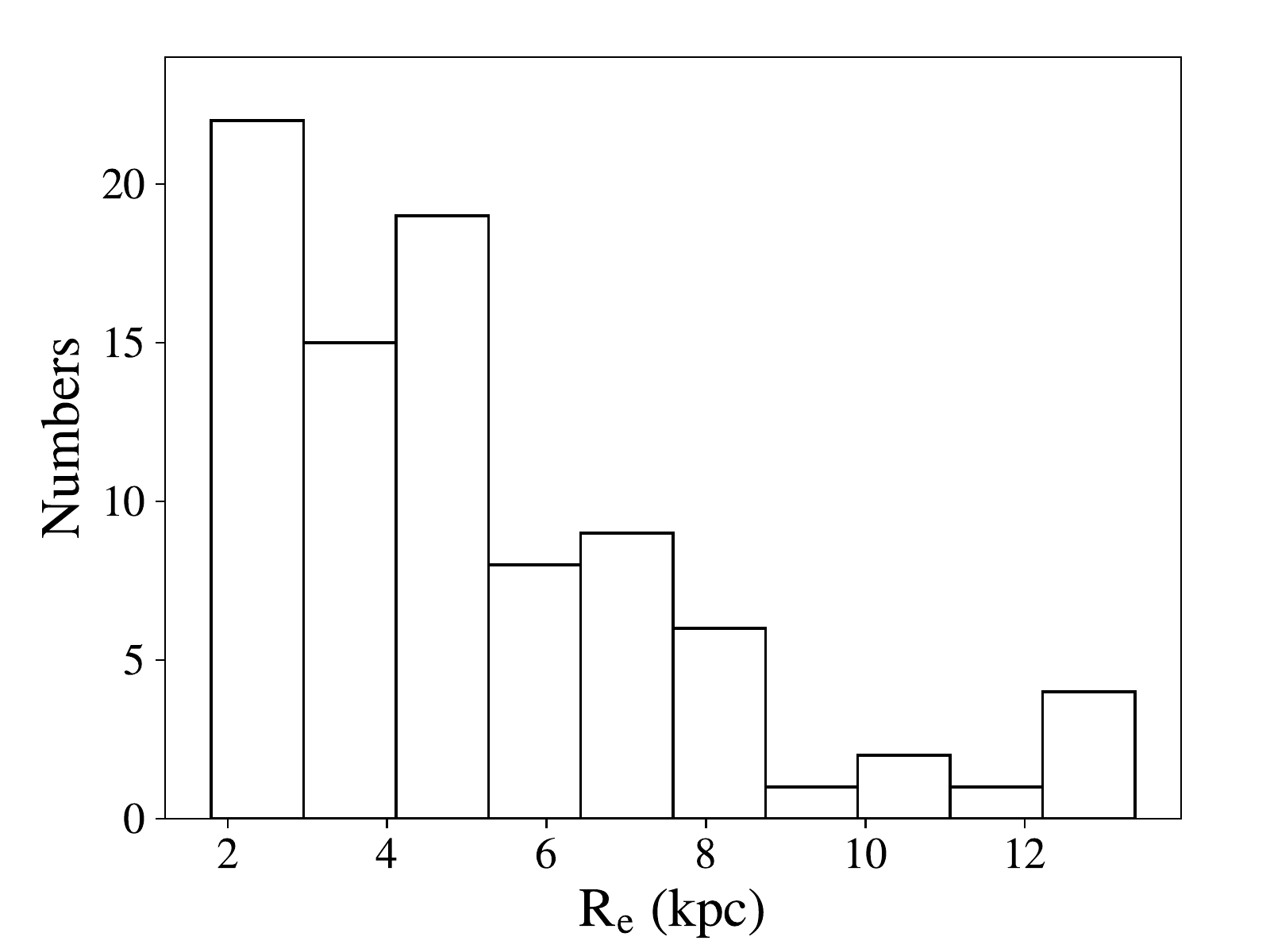}
\plotone{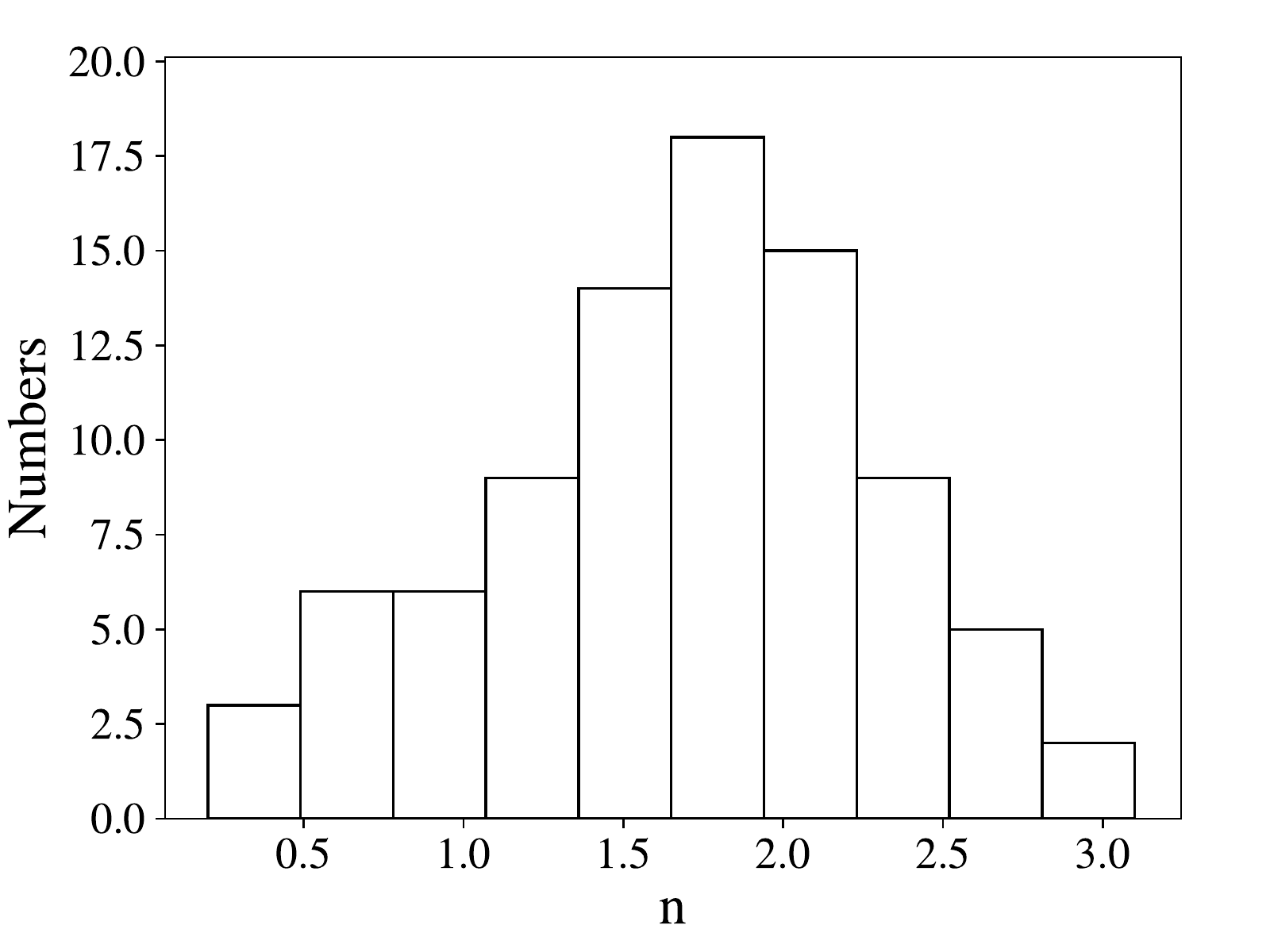}
\caption{The distribution of the S\'ersic parameters of the quasar host galaxies. 
		{\it{Upper Panel}}: The distribution of the half-light radius $R_e$.
		{\it{Lower Panel}}: The distribution of the S\'ersic index $n$.
		In our sample, $n$ spans from $\sim0.5$ (disk-like) to $\sim3$ (bulge-like).
		Most galaxies have $n<2$, indicating 
		that the majority of them are disk-dominated.}
\label{fig:n}
\end{figure}

\begin{figure}
\centering
\includegraphics[trim={1cm 0 0 0},width=0.47\textwidth]{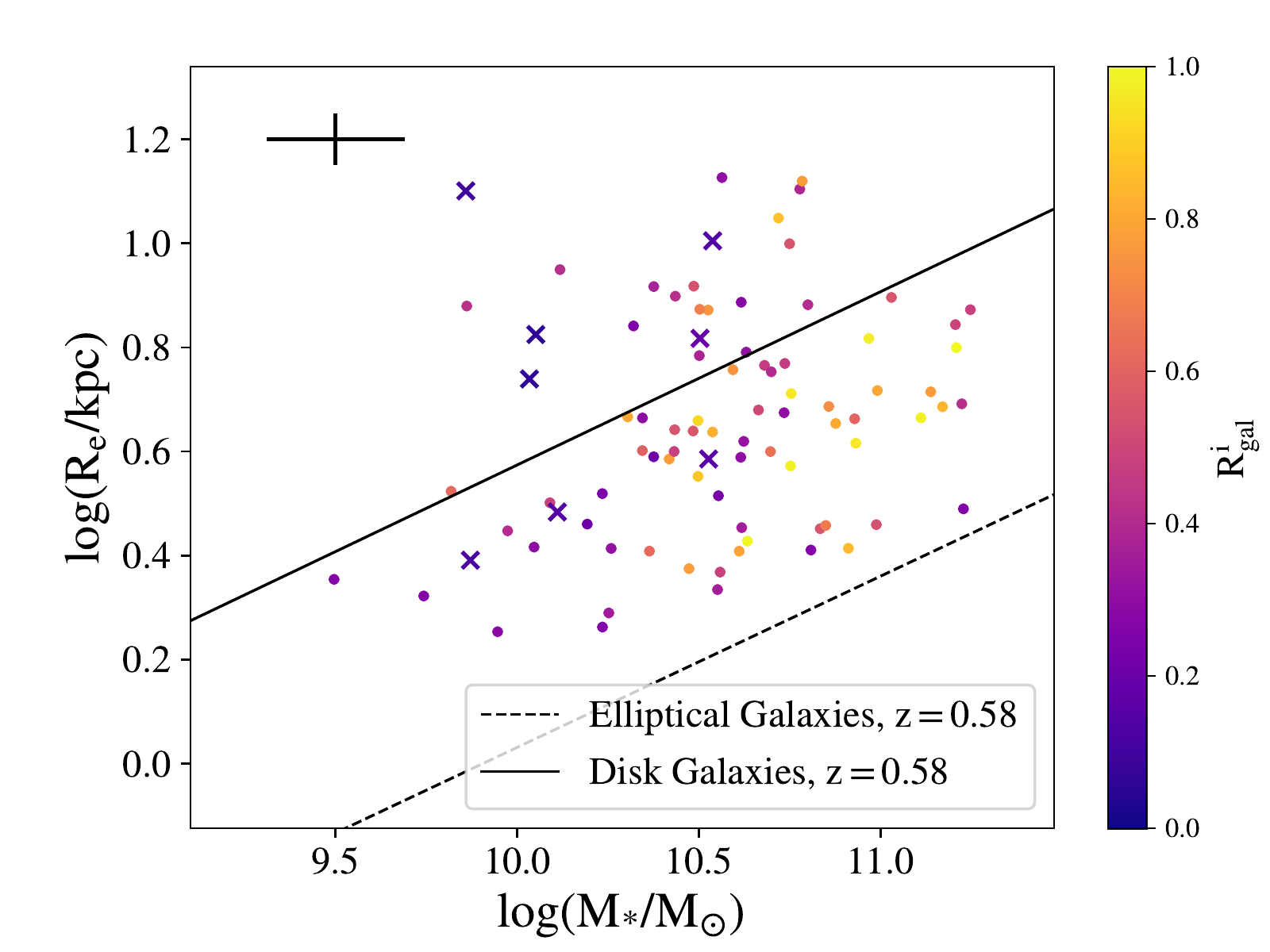}
\caption{The $M_*-R_e$ relation of the quasar host galaxies. 
		The crosses represent the quasars with $R^i_\text{gal}<0.2$,
		and the black dots shows the quasars with $R^i_\text{gal}>0.2$.
		The dashed and solid lines represent the relations 
		for disk and elliptical galaxies at $z=0.58$ (median redshift of our sample), 
		based on \citet{lange16} and \citet{vdw14} (see text). 
		The quasar host galaxies in our sample are consistent with late-type galaxies.
		}
\label{fig:rems}
\end{figure}

\subsection{S\'ersic Parameters}
Figure \ref{fig:n} shows the distribution of the half-light radius $R_e$
and S\'ersic index $n$ of the host galaxies. 
The $R_e$ values span a wide range from $\sim2$ to $\sim13$ kpc, with an average 
of $\sim5$ kpc.
At $z=0.8$, a half-light radius of 2 kpc corresponds to an angular size of $0\farcs27$, 
or $0\farcs54$ in diameter, which can be marginally resolved by our $i$-band images.
The $n$ values span from $\sim0.5$ to $\sim3$. 
Figure \ref{fig:rems} shows the relation between $M_{*}$ and $R_e$ of the quasar host galaxies. 
For comparison, we include in Figure \ref{fig:rems} the $M_*-R_e$ relations 
for disk and elliptical galaxies at $z=0.58$ (the median redshift of our quasar sample).
These relations are estimated as follows. 
We start with the $M_*-R_e$ relations for local disk and elliptical galaxies at $z<0.06$ 
from \citet{lange16}. Galaxies at higher redshift tend to have smaller 
sizes \citep[e.g.,][]{Trujillo07,vdw08,vdw14}. For example, 
\citet{vdw14} reported that, at $0<z<3$, the average radius evolves with redshift as $r \propto (1+z)^{-0.75}$ for late-type galaxies and $r \propto (1+z)^{-1.48}$ for early-type galaxies.
We take into account this size evolution and find that from $z=0$ to $z=0.58$,
the radius of late-type galaxies decreases by $\sim0.15$ dex and the radius of early-type 
galaxies decreases by $\sim0.3$ dex. The relations are plotted in Figure \ref{fig:rems}.
Figure \ref{fig:rems} indicates that the $M_*-R_e$ relation in our quasar host galaxies 
is consistent with the relation for late-type galaxies.
In summary, the distributions of S\'ersic index and half-light radius indicate that 
most of our quasar host galaxies are disk-dominated.

\section{Discussion} \label{discussion}

\subsection{Sample Bias} \label{bias}
Most SDSS-RM quasars are drawn from the SDSS quasar catalog, 
with a small fraction ($\sim5\%$) of quasars discovered by
Panoramic Survey Telescope and Rapid Response System (Pan-STARRS)
Medium Deep Field survey \citep{chambers16} and the DEEP2 survey \citep{newman13}. 
The final sample is flux-limited ($i<21.7$ mag) and objects with fiber collisions are removed.
Since most sources in this sample are SDSS quasars, the completeness 
of our sample depends strongly on the completeness of the SDSS quasar selection. 
The completeness of SDSS quasar catalogs has been investigated in many previous papers. 
For example, \citet{van05} found that the completeness in SDSS-I is about 89\% 
to its limiting magnitude. Quasars from Pan-STARRS and DEEP2 survey 
fill in quasars missed by SDSS, and 
thus these quasars may further increase the sample completeness. 
In addition, \citet{shen15} found that the number of quasars in this sample 
is consistent with the number predicted by the quasar luminosity function.
All the above indicate that the SDSS-RM quasar sample is fairly complete 
to its flux limit. 

Another bias is introduced by the ``Successful Fitting Criteria" 
in the image decomposition process.
The image decomposition is likely to fail for faint host galaxies,
which means that we might miss some faint galaxies.
Since only 8 out of 103 objects are rejected in this step,
and the ``success rate" is larger than 80\% in all redshift bins,
this bias does not affect our results.

\subsection{Massive Star Forming Quasar Host Galaxies}

Our results are consistent with many previous results based on image decomposition. 
For example, \citet{mat14} studied $z<0.6$ SDSS quasars using the 
$g$ and $i$ bands (shifted to $z=0.3$; denoted by $^{0.3}g$ and $^{0.3}i$) 
to construct the CMDs of quasar host galaxies and normal galaxies.
To directly compare to their results, we select a subsample of our quasars 
at $0.35<z<0.55$ for which $^{0.3}g$ and $^{0.3}i$ are available from our spectra.  
This subsample has $^{0.3}(g-i)=1.35 \pm 0.37$ and absolute magnitude $<^{0.3}M_i>=-21.4$, 
which are similar to the results of \citet{mat14}.


\citet{xu15a} studied $\sim 200$ AGNs at $z<2$ selected from the 24-$\mu$m infrared emission, 
and measured their SFRs by SED fitting. 
In a subsequent study, \citet{xu15b} concluded that these AGN host galaxies typically 
have specific SFR consistent with the star-forming main-sequence galaxies. 
Compared to our work, \citet{xu15a} and \citet{xu15b} applied a different method to 
measure the SFRs of galaxies, but reached a similar result.

We also compare our results with \citet{mat15}, 
who studied $z<1$ SDSS-RM quasars by decomposing the co-added spectra.
They found that the rest-frame $u-r$ colors of the host galaxies are between 0.5 and 2.5
with a median value of $\sim2.0$. These galaxies are preferencially located in the ``green valley",
indicating relatively old stellar populations ($\sim 1.0$ Gyr).
The spectral decomposition method in \citet{mat15} assumed single stellar populations (SSP).
We measure the $u-r$ colors of our host galaxies at $0.2<z<0.5$, 
where the rest-frame $u$ and $r$ are covered by the BOSS spectra.
The $u-r$ colors are roughly between 0.5 and 2.0, with the median value of $\sim1.4$.
For comparison, the blue cloud of inactive galaxies has $u-r\sim1.2$ in our control sample,
thus our quasar host galaxies have similar colors to the blue cloud.
The reason for the discrepancy between our results and the \citet{mat15} results is not clear.

\begin{figure}
\epsscale{1.2}
\plotone{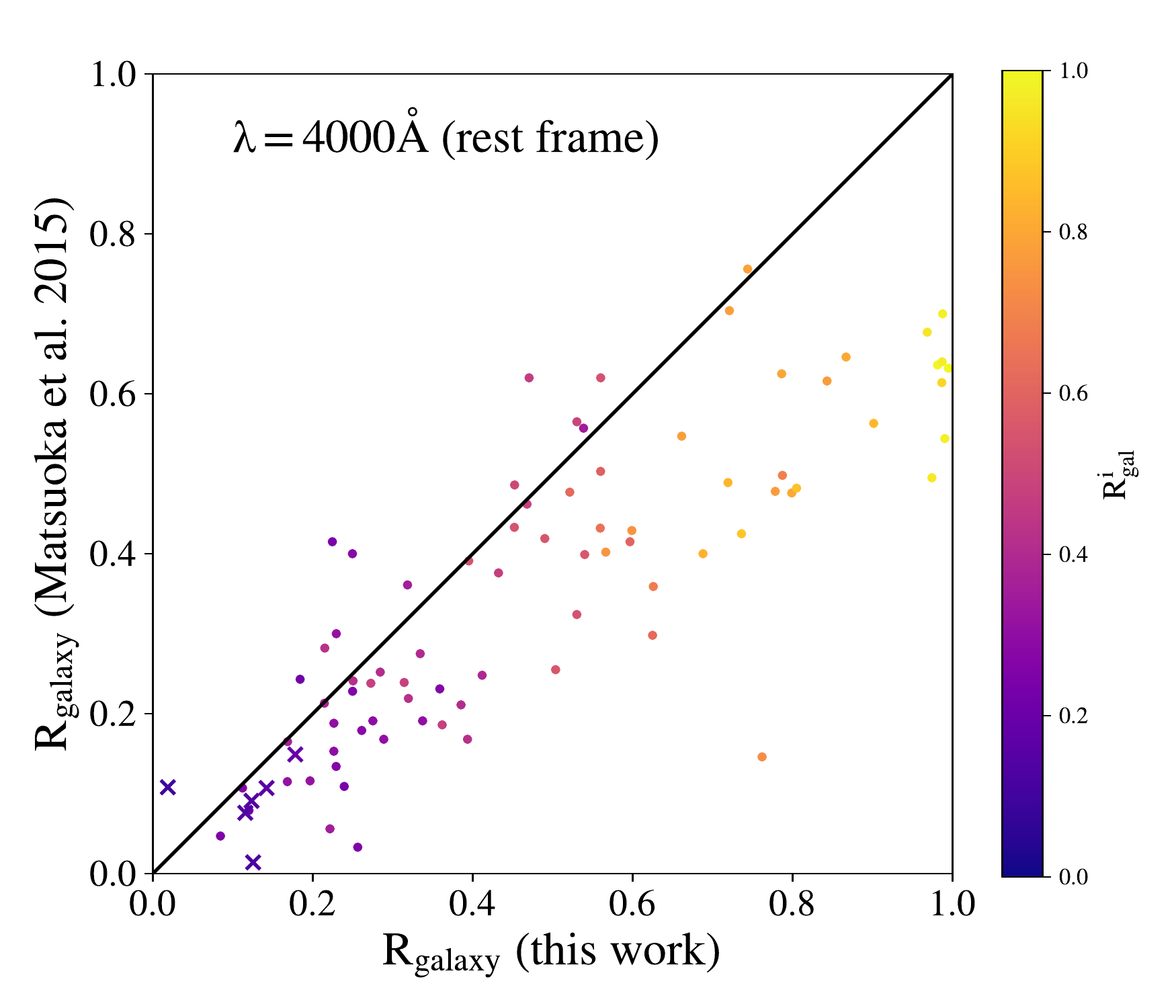}
\plotone{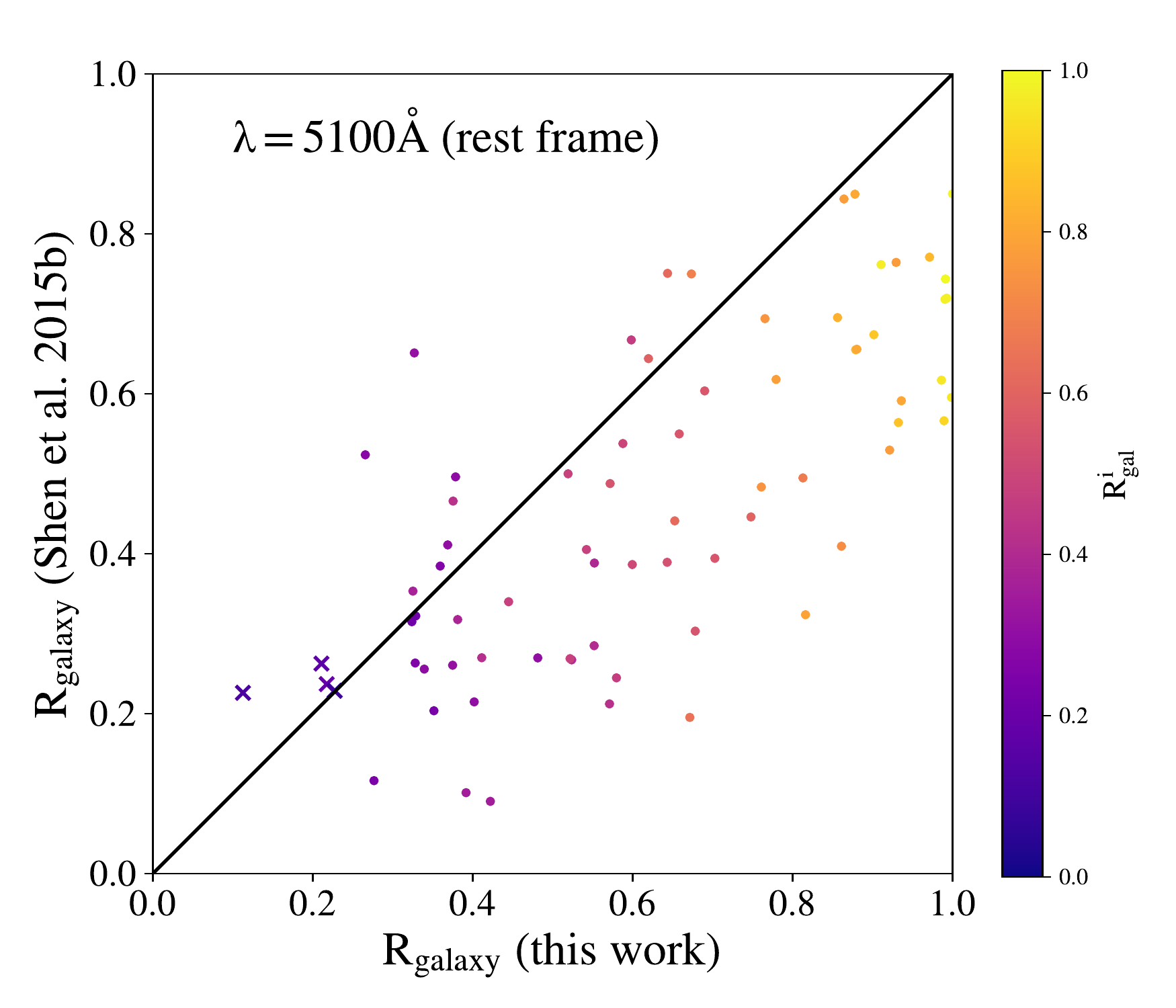}
\caption{
{\it{Upper Panel}}:
The fraction of the host galaxy flux in the total flux (host fraction) at the rest-frame 4000{\AA}
from this work and \citet{mat15}.
The crosses represent the quasar host galaxies with $R^i_\text{gal}<0.2$,
and the black dots represent the galaxies with $R^i_\text{gal}>0.2$.
There is a positive correlation between our results and \citet{mat15},
while our results are systematically larger by $\sim0.15$.
The difference increases with the host fraction.
{\it{Lower Panel}}: The host fraction at the rest-frame 5100{\AA} from this work and \citet{shen15b}.
Our results are larger by $\sim0.14$ compared to \citet{shen15b}.
These results indicate that the host galaxy flux provided by image decomposition methods 
might be systematically larger than the results from spectra decompositions.}
\label{fig:fh}
\end{figure}

The low-redshift $(z<1)$ SDSS-RM quasar sample has been
analyzed using spectra decomposition by \citet{mat15} and by \citet{shen15b}. 
So our sample is actually a subset of the sample 
used by these two studies. We use the same sample to investigate the different results 
from image and spectral decomposition.
\citet{mat15} provided the fraction of the host galaxy in the total flux (host fraction)
at the rest-frame 4000 {\AA}.
For objects that are successfully fitted in both their and our work,
we compare the host fraction provided by the two different methods.
There is a positive correlation between the two results,
but our result is systematically larger.
The median of the difference between the host fraction
of our work and \citet{mat15} is $0.15$.
The difference becomes larger with increasing host fraction.
\citet{shen15b} decomposed the SDSS-RM quasar spectra
using principal component analysis method, and provided 
the host fraction at the rest-frame 5100 {\AA}.
We compare their result with our host fraction at the rest-frame 5100 {\AA} 
and find that our result is larger by $\sim0.14$.
Figure \ref{fig:fh} shows the two comparisons.
This difference may explain the bluer colors of our host galaxies compared to \citet{mat15}.
In short, our result is consist with most previous studies using image decomposition, but 
systematically larger than the results from spectral decomposition. This may indicate that
the galaxy flux estimated from image decomposition is generally larger
than the flux from spectral decomposition.

\subsection{$M_*-M_{\text{BH}}$ Relation of Quasar Host Galaxies}

Previous studies have 
reported different results of the $M_*-M_{\text{BH}}$ relation for quasar host galaxies.
For example, \citet{mat14} and \citet{mat15} showed a positive correlation between 
$M_*$ and $M_{\text{BH}}$, while \citet{falomo14} suggested no correlation.

As shown in Figure \ref{fig:mbh}, there is clearly a positive $M_*-M_{\text{BH}}$ correlation 
in our sample. The best-fit of the $M_*-M_{\text{BH}}$ relation is
\begin{multline}
\text{log}(\frac{M_\text{BH}}{1.18 \times 10^8M_\odot})= \\
(0.55 \pm 0.13)\times\text{log}(\frac{M_*}{3.53\times 10^{10}M_\odot}) 
 + (0.00\pm 0.05).
\end{multline}
The errors of the fitting parameters are estimated by Bootstrap.
This relation is shallower than the $M_\text{Bulge}-M_{\text{BH}}$ relation for local galaxies.
The slope of the $M_\text{BH}-M_\text{Bulge}$ relation for local galaxies
is $1.16 \pm 0.08$ \citep{kh13}, which is $3.9\sigma$ larger than our result.
A shallow relation was also reported in \citet{mat14} and \citet{mat15}.
If this shallow relation is physical, it may indicate that the growth of the SMBH mass
and the stellar mass in quasars are complex processes and are not synchronized.
However, previous studies have suggested that selection biases 
can influence the observed $M_*-M_{\text{BH}}$ relation
\citep[e.g.,][]{schulze11,degraf15,shankar16}.
On one hand, the SDSS-RM quasar sample is flux-limited,
some low-luminosity (thus low-SMBH-mass) objects might be missed,
especially at high redshift.
On the other hand, the galaxy sample 
used to calibrate the local $M_\text{Bulge}-M_{\text{BH}}$ relation
might also be biased,
because a significant fraction of galaxies are selected to have 
dynamically measured SMBH masses, which requires that
 the black hole sphere of influence must be resolved \citep{shankar16}.
 The errors of single-epoch SMBH mass measurements
 may also have significant influence on the observed $M_*-M_{\text{BH}}$ relation
 \citep[e.g.,][]{lauer07,shen10}.
 It is difficult to tell whether the discrepancies between quasar host galaxies and local galaxies
 shown in our results are physical.
A detailed discussion about the influence of selection effects on the 
shallowness of the observed $M_*-M_{\text{BH}}$ relation can be found in \citet{shen15b}.

There are some other sources of systematic errors.
First, the stellar masses in our results are calculated using the broad-band flux
that is measured according to the co-added SDSS-RM spectra.
The diameter of the spectrograph fiber is $2''$.
Given the wide range of the half-light radius in our sample,
we may have missed some flux for some large, low-redshift galaxies,
and thus underestimated their stellar masses.
Since the measurement of the S\'ersic parameters are not accurate,
especially for the S\'ersic index,
it is difficult to correct this systematic error.
Second, the stellar masses in our results include both disk and bulge components.
The stellar masses should be regarded as upper limits when investigating the 
 $M_\text{Bulge}-M_{\text{BH}}$ relation of the quasar hosts.
 Including disk component in the stellar mass can introduce
 significant scatters to the $M_*-M_{\text{BH}}$ relation.
For example, \citet{falomo14} performed image decomposition on $z<0.5$ quasars.
Different from our approach, they decomposed the host galaxies into bulge and disk components.
Their sample shows no correlation between $M_*$ and $M_{\text{BH}}$,
while there is a significant correlation between $M_\text{Bulge}$ and $M_{\text{BH}}$.
Our morphology analysis indicates that the host galaxies in our sample have prominent disk components,
and may be affected by this systematic error.

\subsection{Morphology of Quasar Host Galaxies} \label{morph}
Figure \ref{fig:rems} shows the $M_*-R_e$ relation of the quasar host galaxies in our sample. 
It suggests that the quasar host galaxies are more consistent with late-type galaxies 
rather than early-type galaxies.
The distribution of the S\'ersic index also supports this point. 
As shown in Figure \ref{fig:n}, about 70\% galaxies have S\'ersic index $0.5<n<2$ (disk-like). 
These results indicate that a significant fraction of quasar host galaxies are disk-dominated. 
\citet{falomo14} reached a similar conclusion for $z<0.6$ quasars.

Morphology of quasar host galaxies can constrain the evolution model of quasars. 
In major merger models, the quasar host galaxies are expected to be either ellipticals or 
interacting galaxies, 
while secular evolution can produce disk-like AGN hosts. 
Our result suggests that a significant fraction of quasars with $-25<M_g<-17$
 at $0.2<z<0.8$ are more likely to form by secular evolution. 
This result is consistent with previous studies \citep[e.g.,][]{cisternas11,villforth17}, 
which showed that most low-redshift $(z<1)$ AGN hosts did not exhibit signs of mergers.

\section{Summary} \label{sum}

We have presented the properties of the host galaxies of 103 $z<0.8$ quasars 
in the SDSS-RM field. We combined images taken by CFHT/MegaCam, and obtained 
deep co-added images with $5\sigma$ depth of $\sim 26$ mag in the $i$ band.
Each quasar image is decomposed into a PSF and a S\'ersic profile, representing the AGN 
and the galaxy component.
A total of 95 out of 103 quasars were successfully decomposed.
The systematic error of the galaxy magnitudes is $\sim0.3$ mag, which is significantly
smaller than the errors in most previous ground-based studies.
Our main results are:
\begin{enumerate}
\item{The quasar host galaxies are more massive ($M_*\sim10^{10.5}M_{\odot}$) than
inactive galaxies with the same redshifts. 
They have rest frame $u-g\sim 0.7$ which is similar to star-forming galaxies.}
\item{The flux from host galaxies is comparable to quasar flux.
	The typical value of the AGN-to-galaxy flux ratio is $\sim 2.5$ in the rest-frame $u$ band 
	and $\sim 2$ in the rest-frame $g$ band.
	  These ratios show little redshift dependence at $0.2<z<0.8$.}
\item{The $M_*-M_\text{BH}$ relation for the quasar host galaxies in our sample is shallower than the local
		$M_\text{Bulge}-M_\text{BH}$ relation. This discrepancy may be physical or originate from complex biases.}
\item{The distribution of the S\'ersic indices and the $M_*-R_e$ relation in our sample 
	 indicate that these quasar hosts are dominated by disk-like galaxies.}
\end{enumerate}

Our study demonstrates that deep ground-based imaging data with excellent PSF 
are able to provide reliable estimate of broad-band flux and morphological information for
low-redshift quasar host galaxies. In this study, we only have two band data, $g$ and $i$.
The large upcoming multi-wavelength sky surveys with great depth and seeing, such as
the Hyper-Suprime Cam survey \citep{aih17} and the Large Synoptic Survey 
Telescope \citep{lsst}
will largely expand the quasar sample that is suitable for the image decomposition method
and provide more solid conclusions.
In addition, future results of the SDSS-RM project will provide more accurate measurements
of the BH masses of these quasars, which is crucial for drawing more reliable conclusions
about the growth history of SMBH and stellar mass in these quasar host galaxies.

\acknowledgments

We thank the referee for many useful comments that have significantly improved this work.
We acknowledge support from the National Key R\&D Program of China (2016YFA0400703) and 
from the National Science Foundation of China (11533001).
YS acknowledges support from an Alfred P. Sloan Research Fellowship and NSF grant AST-1715579.
PBH is supported by NSERC.
LCH was supported by the National Key R\&D Program of China (2016YFA0400702) 
and the National Science Foundation of China (11473002, 11721303).
KH acknowledges support from STFC grant ST/M001296/1.

Funding for SDSS-III has been provided by the Alfred P. Sloan Foundation, the Participating Institutions, the National Science Foundation, and the U.S. Department of Energy Office of Science. The SDSS-III web site is http://www.sdss3.org/.
SDSS-III is managed by the Astrophysical Research Consortium for the Participating Institutions of the SDSS-III Collaboration including the University of Arizona, the Brazilian Participation Group, Brookhaven National Laboratory, Carnegie Mellon University, University of Florida, the French Participation Group, the German Participation Group, Harvard University, the Instituto de Astrofisica de Canarias, the Michigan State/Notre Dame/JINA Participation Group, Johns Hopkins University, Lawrence Berkeley National Laboratory, Max Planck Institute for Astrophysics, Max Planck Institute for Extraterrestrial Physics, New Mexico State University, New York University, Ohio State University, Pennsylvania State University, University of Portsmouth, Princeton University, the Spanish Participation Group, University of Tokyo, University of Utah, Vanderbilt University, University of Virginia, University of Washington, and Yale University.

Based on observations obtained at the Canada-France-Hawaii Telescope (CFHT) which is operated by the National Research Council of Canada, the Institut National des Sciences de l'Univers of the Centre National de la Recherche Scientifique of France, and the University of Hawaii. The authors recognize and acknowledge the very significant cultural role and reverence that the summit of Maunakea has always had within the indigenous Hawaiian community.  We are most fortunate to have the opportunity to conduct observations from this mountain.

\facilities{CFHT (MegaCam), SDSS}

\appendix

\section{Modeling PSF using PSFEx} \label{ap:psf}
\begin{figure}[h]
\epsscale{1}
\plotone{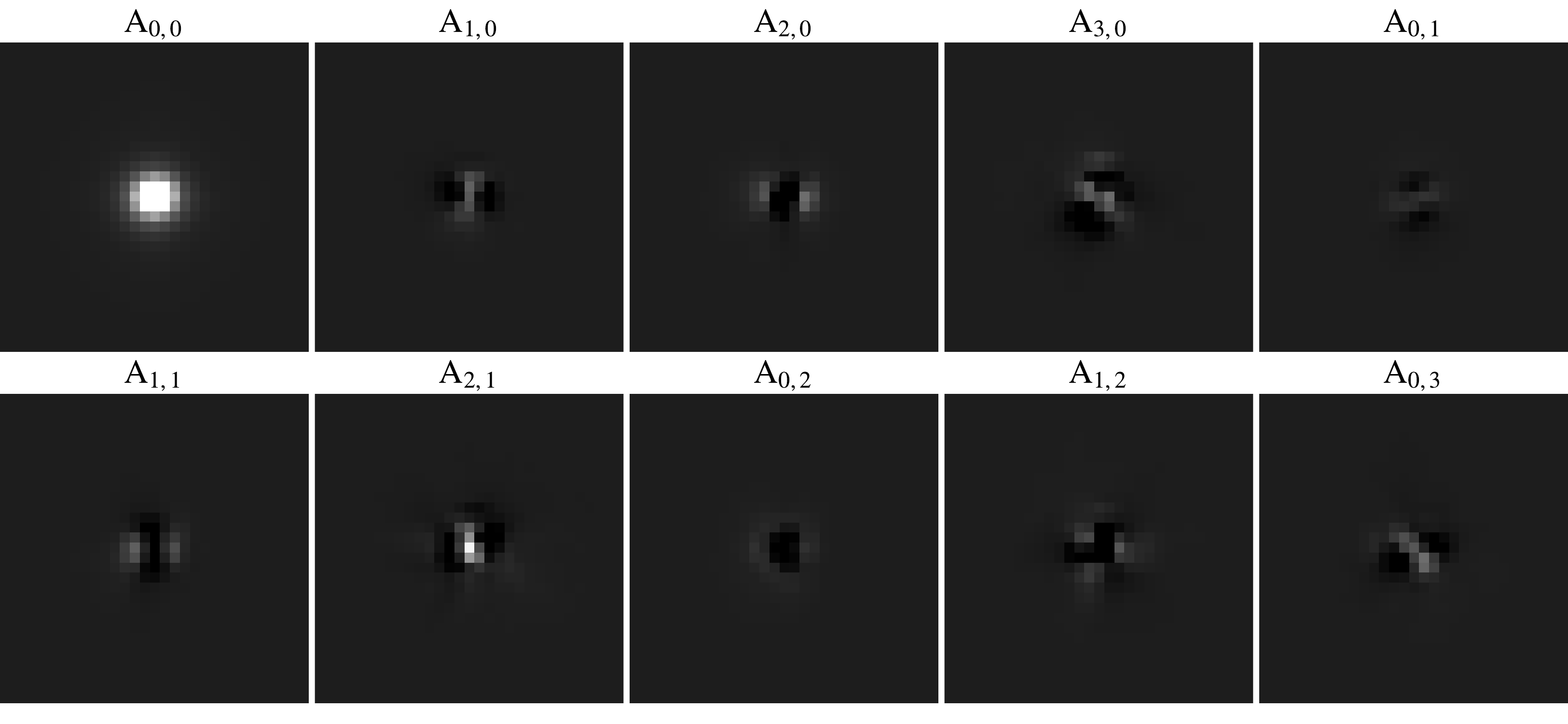}
\caption{The PSF model of $i$ band image of A pointing generated by {\texttt{PSFEx}}. See text for the meaning of $A_{m,n}$.
The images are shown in logarithmic scale. The $A_{0,0}$ component is the dominate component (the central pixels are saturated
to show the details in the other components).}
\label{fig:psfmodel}
\end{figure}

\begin{figure}[h]
\epsscale{0.7}
\plotone{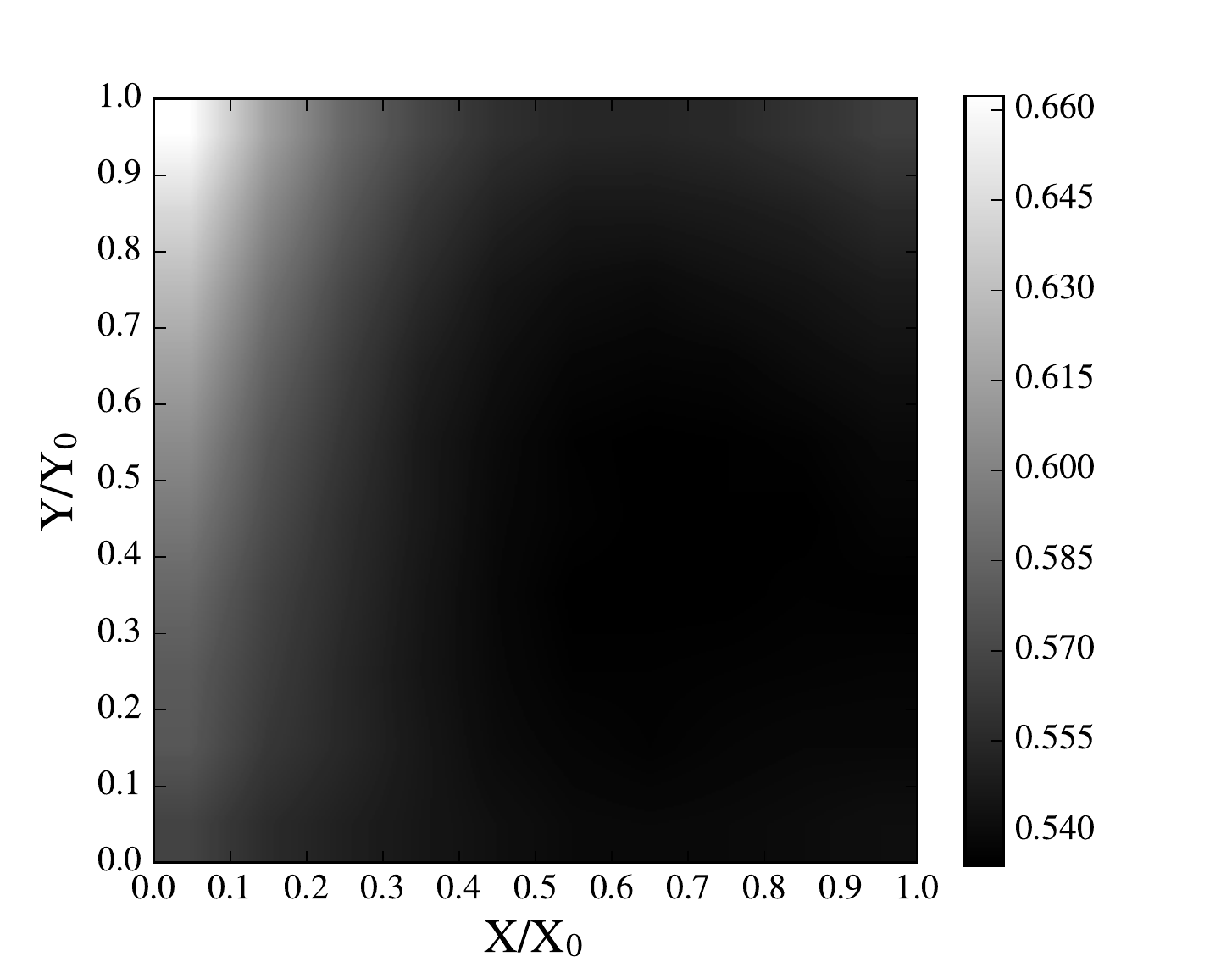}
\caption{The variation of PSF FWHM across an image. This figure takes the $i$ band image of A pointing as an example. 
$X_0$ and $Y_0$ are the size of the image (in pixel). The FWHM of the PSF is 
expressed in arcsec. The variation of PSF FWHM is $\lesssim 10\%$ in the whole image.}
\label{fig:psfvar}
\end{figure}

{\texttt{PSFEx}} models a PSF using a polynomial function:
\begin{equation} \label{eq:PSF}
\text{PSF}(x,y,i,j)=\sum_{m+n \le N}{A_{m,n}(i,j)x^my^n}
\end{equation}
where $x,y$ are the position on the detector, $i,j$ mark the pixel in the PSF model, and $N$ is the degree of the polynomial function.
{\texttt{PSFEx}} provides various choices of the function $A_{m,n}(i,j)$, including pixel-based (i.e., the value of each pixel 
is a free parameter and can change independently),
and some commonly-used analytical functions (e.g., Gaussian, Moffat). 
When generating the PSF model, {\texttt{PSFEx}} selects bright, unsaturated, point-like objects based on their 
flux and half-flux radius, and fits the PSF model in Equation \ref{eq:PSF} by $\chi^2$-minimization.
In this study we set $N=3$ and model the PSF in a pixel-based style.
Using the $i$ band image of Pointing A as an example, Figure \ref{fig:psfmodel} shows the images of $A_{m,n}(i,j)$, and
Figure \ref{fig:psfvar} shows the variation of the PSF FWHM across the detector (including all 36 CCD chips). 
According to Figure \ref{fig:psfvar}, that the variation of PSF FWHM across a image is $\lesssim 10\%$.

\begin{thebibliography}{}

\bibitem[Aihara et al.(2017)]{aih17} Aihara, H., Arimoto, N., Armstrong, R., et al.\ 2017, arXiv:1704.05858
\bibitem[Annis et al.(2014)]{annis14} Annis, J., Soares-Santos, M., Strauss, M.~A., et al.\ 2014, \apj, 794, 120
\bibitem[Aird et al.(2012)]{aird12} Aird, J., Coil, A.~L., Moustakas, J., et al.\ 2012, \apj, 746, 90
\bibitem[Aune et al.(2003)]{aune03} Aune, S., Boulade, O., Charlot, X., et al.\ 2003, \procspie, 4841, 513
\bibitem[Bahcall et al.(1997)]{bahcall97} Bahcall, J.~N., Kirhakos, S., Saxe, D.~H., \& Schneider, D.~P.\ 1997, \apj, 479, 642 
\bibitem[Beers et al.(1990)]{beer90} Beers, T.~C., Flynn, K., \& Gebhardt, K.\ 1990, \aj, 100, 32
\bibitem[Bell et al.(2003)]{bell03} Bell, E.~F., McIntosh, D.~H., Katz, N., \& Weinberg, M.~D.\ 2003, \apjs, 149, 289
\bibitem[Bettoni et al.(2015)]{bettoni15} Bettoni, D., Falomo, R., Kotilainen, J.~K., Karhunen, K., \& Uslenghi, M.\ 2015, \mnras, 454, 4103
\bibitem[Bertin \& Arnouts(1996)]{bertin96} Bertin, E., \& Arnouts, S.\ 1996, \aaps, 117, 393 
\bibitem[Bertin et al.(2002)]{bert02} Bertin, E., Mellier, Y., Radovich, M., et al.\ 2002, Astronomical Data Analysis Software and Systems XI, 281, 228 
\bibitem[Bertin(2011)]{bertin11} Bertin, E.\ 2011, Astronomical Data Analysis Software and Systems XX, 442, 435 
\bibitem[Brown et al.(2014)]{brown14} Brown, M.~J.~I., Moustakas, J., Smith, J.-D.~T., et al.\ 2014, \apjs, 212, 18 
\bibitem[Cardelli et al.(1989)]{Cardelli89} Cardelli, J.~A., Clayton, G.~C., \& Mathis, J.~S.\ 1989, \apj, 345, 245
\bibitem[Chambers et al.(2016)]{chambers16} Chambers, K.~C., Magnier, E.~A., Metcalfe, N., et al.\ 2016, arXiv:1612.05560
\bibitem[Cisternas et al.(2011)]{cisternas11} Cisternas, M., Jahnke, K., Inskip, K.~J., et al.\ 2011, \apj, 726, 57 
\bibitem[Cicone et al.(2014)]{cicone14} Cicone, C., Maiolino, R., Sturm, E., et al.\ 2014, \aap, 562, A21 
\bibitem[Croton et al.(2006)]{croton06} Croton, D.~J., Springel, V., White, S.~D.~M., et al.\ 2006, \mnras, 365, 11
\bibitem[DeGraf et al.(2015)]{degraf15} DeGraf, C., Di Matteo, T., Treu, T., et al.\ 2015, \mnras, 454, 913
\bibitem[Di Matteo et al.(2005)]{DM05} Di Matteo, T., Springel, V., \& Hernquist, L.\ 2005, \nat, 433, 604 
\bibitem[Eisenstein et al.(2011)]{eisenstein11} Eisenstein, D.~J., Weinberg, D.~H., Agol, E., et al.\ 2011, \aj, 142, 72
\bibitem[Fabian(2012)]{fabian12} Fabian, A.~C.\ 2012, \araa, 50, 455
\bibitem[Falomo et al.(2014)]{falomo14} Falomo, R., Bettoni, D., Karhunen, K., Kotilainen, J.~K., \& Uslenghi, M.\ 2014, \mnras, 440, 476 
\bibitem[Georgakakis et al.(2008)]{Georgakakis08} Georgakakis, A., Nandra, K., Yan, R., et al.\ 2008, \mnras, 385, 2049 
\bibitem[Greene \& Ho(2005)]{greene05} Greene, J.~E., \& Ho, L.~C.\ 2005, \apj, 630, 122
\bibitem[Gunn et al.(2006)]{gun06} Gunn, J.~E., Siegmund, W.~A., Mannery, E.~J., et al.\ 2006, \aj, 131, 2332 
\bibitem[Hickox et al.(2009)]{Hickox09} Hickox, R.~C., Jones, C., Forman, W.~R., et al.\ 2009, \apj, 696, 891 
\bibitem[Hopkins et al.(2006)]{hopkins06} Hopkins, P.~F., Hernquist, L., Cox, T.~J., et al.\ 2006, \apjs, 163, 1 
\bibitem[Jahnke et al.(2004)]{jahnke04} Jahnke, K., S{\'a}nchez, S.~F., Wisotzki, L., et al.\ 2004, \apj, 614, 568
\bibitem[Jiang et al.(2014)]{jiang14} Jiang, L., Fan, X., Bian, F., et al.\ 2014, \apjs, 213, 12 
\bibitem[Kauffmann et al.(2003)]{Kauffmann03} Kauffmann, G., Heckman, T.~M., Tremonti, C., et al.\ 2003, \mnras, 346, 1055
\bibitem[Kim et al.(2008)]{kim08} Kim, M., Ho, L.~C., Peng, C.~Y., et al.\ 2008, \apj, 687, 767-827
\bibitem[King \& Pounds(2015)]{king15} King, A., \& Pounds, K.\ 2015, \araa, 53, 115
\bibitem[Kirhakos et al.(1999)]{Kirhakos99} Kirhakos, S., Bahcall, J.~N., Schneider, D.~P., \& Kristian, J.\ 1999, \apj, 520, 67 
\bibitem[Kormendy \& Ho(2013)]{kh13} Kormendy, J., \& Ho, L.~C.\ 2013, \araa, 51, 511 
\bibitem[Lange et al.(2016)]{lange16} Lange, R., Moffett, A.~J., Driver, S.~P., et al.\ 2016, \mnras, 462, 1470 
\bibitem[Lauer et al.(2007)]{lauer07} Lauer, T.~R., Tremaine, S., Richstone, D., \& Faber, S.~M.\ 2007, \apj, 670, 249
\bibitem[LSST Dark Energy Science Collaboration(2012)]{lsst} LSST Dark Energy Science Collaboration 2012, arXiv:1211.0310 
\bibitem[Matsuoka et al.(2014)]{mat14} Matsuoka, Y., Strauss, M.~A., Price, 
	T.~N., III, \& DiDonato, M.~S.\ 2014, \apj, 780, 162 
\bibitem[Matsuoka et al.(2015)]{mat15} Matsuoka, Y., Strauss, M.~A., Shen, 
	Y., et al.\ 2015, \apj, 811, 91 
\bibitem[McLure et al.(1999)]{mclure99} McLure, R.~J., Kukula, M.~J., Dunlop, J.~S., et al.\ 1999, \mnras, 308, 377
\bibitem[Muzzin et al.(2013)]{muzzin13} Muzzin, A., Marchesini, D., Stefanon, M., et al.\ 2013, \apjs, 206, 8
\bibitem[Nandra et al.(2007)]{nandra07} Nandra, K., Georgakakis, A., Willmer, C.~N.~A., et al.\ 2007, \apjl, 660, L11
\bibitem[Newman et al.(2013)]{newman13} Newman, J.~A., Cooper, M.~C., Davis, M., et al.\ 2013, \apjs, 208, 5
\bibitem[Oke \& Gunn(1983)]{oke83} Oke, J.~B., \& Gunn, J.~E.\ 1983, \apj, 266, 713
\bibitem[P{\^a}ris et al.(2017)]{paris17} P{\^a}ris, I., Petitjean, P., Ross, N.~P., et al.\ 2017, \aap, 597, A79 
\bibitem[Salom{\'e} et al.(2015)]{salome15} Salom{\'e}, Q., Salom{\'e}, P., \& Combes, F.\ 2015, \aap, 574, A34
\bibitem[Schlegel et al.(1998)]{Schlegel98} Schlegel, D.~J., Finkbeiner, D.~P., \& Davis, M.\ 1998, \apj, 500, 525
\bibitem[Schulze \& Wisotzki(2011)]{schulze11} Schulze, A., \& Wisotzki, L.\ 2011, \aap, 535, A87
\bibitem[Sersic(1968)]{Sersic68} Sersic, J.~L.\ 1968, Cordoba, Argentina: Observatorio Astronomico
\bibitem[Shankar et al.(2016)]{shankar16} Shankar, F., Bernardi, M., Sheth, R.~K., et al.\ 2016, \mnras, 460, 3119
\bibitem[Shen \& Kelly(2010)]{shen10} Shen, Y., \& Kelly, B.~C.\ 2010, \apj, 713, 41
\bibitem[Shen et al.(2015a)]{shen15} Shen, Y., Brandt, W.~N., Dawson, K.~S., 
	et al.\ 2015, \apjs, 216, 4 
\bibitem[Shen et al.(2015b)]{shen15b} Shen, Y., Greene, J.~E., Ho, L.~C., et al.\ 2015, \apj, 805, 96
\bibitem[Silverman et al.(2008)]{Silverman08} Silverman, J.~D., Green, P.~J., Barkhouse, W.~A., et al.\ 2008, \apj, 679, 118-139
\bibitem[Smee et al.(2013)]{smee13} Smee, S.~A., Gunn, J.~E., Uomoto, A., et al.\ 2013, \aj, 146, 32
\bibitem[Spacek et al.(2016)]{spacek16} Spacek, A., Scannapieco, E., Cohen, S., Joshi, B., \& Mauskopf, P.\ 2016, \apj, 819, 128
\bibitem[Springel et al.(2005)]{Springel05} Springel, V., Di Matteo, T., \& Hernquist, L.\ 2005, \mnras, 361, 776 
\bibitem[Trujillo et al.(2007)]{Trujillo07} Trujillo, I., Conselice, C.~J., Bundy, K., et al.\ 2007, \mnras, 382, 109
\bibitem[Trump et al.(2013)]{trump13} Trump, J.~R., Hsu, A.~D., Fang, J.~J., et al.\ 2013, \apj, 763, 133
\bibitem[Tsuzuki et al.(2006)]{tsu06} Tsuzuki, Y., Kawara, K., Yoshii, Y., et al.\ 2006, \apj, 650, 57
\bibitem[van der Wel et al.(2008)]{vdw08} van der Wel, A., Holden, B.~P., Zirm, A.~W., et al.\ 2008, \apj, 688, 48-58
\bibitem[van der Wel et al.(2014)]{vdw14} van der Wel, A., Franx, M., van Dokkum, P.~G., et al.\ 2014, \apj, 788, 28
\bibitem[van Dokkum(2001)]{vdokkum01} van Dokkum, P.~G.\ 2001, \pasp, 113, 1420
\bibitem[Vanden Berk et al.(2005)]{van05} Vanden Berk, D.~E., Schneider, D.~P., Richards, G.~T., et al.\ 2005, \aj, 129, 2047
\bibitem[V{\'e}ron-Cetty et al.(2004)]{vc04} V{\'e}ron-Cetty, M.-P., Joly, M., \& V{\'e}ron, P.\ 2004, \aap, 417, 515 
\bibitem[Vestergaard \& Peterson(2006)]{vp06} Vestergaard, M., \& Peterson, B.~M.\ 2006, \apj, 641, 689 
\bibitem[Villforth et al.(2017)]{villforth17} Villforth, C., Hamilton, T., Pawlik, M.~M., et al.\ 2017, \mnras, 466, 812
\bibitem[Xu et al.(2015a)]{xu15a} Xu, L., Rieke, G.~H., Egami, E., et al.\ 2015, \apjs, 219, 18 
\bibitem[Xu et al.(2015b)]{xu15b} Xu, L., Rieke, G.~H., Egami, E., et al.\ 2015, \apj, 808, 159
\bibitem[York et al.(2000)]{york00} York, D.~G., Adelman, J., Anderson, J.~E., Jr., et al.\ 2000, \aj, 120, 1579
\bibitem[Zakamska et al.(2005)]{zakamska05} Zakamska, N.~L., Schmidt, G.~D., Smith, P.~S., et al.\ 2005, \aj, 129, 1212
\bibitem[Zinn et al.(2013)]{zinn13} Zinn, P.-C., Middelberg, E., Norris, R.~P., \& Dettmar, R.-J.\ 2013, \apj, 774, 66 
\end{thebibliography}
\end{document}